\documentclass[onecolumn, a4paper, oneside, onecolumn, titlepage]{article}

\usepackage{lineno}					
\usepackage{amsmath}				
\usepackage{amsfonts}				
\usepackage{setspace}				
\usepackage[]{natbib}               
\usepackage[nocaption]{optional}	
\usepackage{subfig}					
\usepackage{graphicx}				
\usepackage{booktabs}				
\usepackage{siunitx}				

\usepackage{todonotes}

\usepackage{etoolbox}
\newbool{MyRefNumbers}

\usepackage[%
	lmargin=2.5cm,%
	rmargin=2.5cm,%
	tmargin=3cm,%
	bmargin=3cm%
]{geometry}%

\newcommand*\patchenviroforlineno[1]{%
	\expandafter\let\csname old#1\expandafter\endcsname\csname #1\endcsname %
	\expandafter\let\csname oldend#1\expandafter\endcsname\csname end#1\endcsname %
	\renewenvironment{#1}%
		{\linenomath\csname old#1\endcsname}%
		{\csname oldend#1\endcsname\endlinenomath}%
}%

\newcommand*\patchbothenviroforlineno[1]{%
	\patchenviroforlineno{#1}%
	\patchenviroforlineno{#1*}
}%

\AtBeginDocument{%
	\patchbothenviroforlineno{equation}%
	\patchbothenviroforlineno{align}%
	\patchbothenviroforlineno{flalign}%
	\patchbothenviroforlineno{alignat}%
	\patchbothenviroforlineno{gather}%
	\patchbothenviroforlineno{multline}%
}%

\begin{document}
\begin{titlepage}
	\vspace{\stretch{1}}
	\begin{center}
		{ \huge \bfseries %
			Eco-evolution from deep time to contemporary dynamics: the role of timescales and rate modulators
		} \\ %
		\vspace{\stretch{1}}
		Emanuel A. Fronhofer$^{1}$, Dov Corenblit$^{2,3}$, Jhelam N. Deshpande$^1$, Lynn Govaert$^4$, Philippe Huneman$^5$, Frédérique Viard$^1$, Philippe Jarne$^6$ and Sara Puijalon$^7$
	\end{center}
	\vspace{\stretch{0.25}}
	\begin{enumerate}
		\item ISEM, Universit\'e de Montpellier, CNRS, IRD, EPHE, Montpellier, France
		\item Université Clermont Auvergne, CNRS, GEOLAB – F-63000 Clermont-Ferrand, France
		\item CNRS, Laboratoire écologie fonctionnelle et environnement, Université Paul Sabatier, CNRS, INPT, UPS, F-31062 Toulouse, France
		\item Leibniz Institute of Freshwater Ecology and Inland Fisheries, Müggelseedamm 310, 12587 Berlin, Germany
		\item Institut d’Histoire et de Philosophie des Sciences et des Techniques (CNRS/ Université Paris I Sorbonne)
		\item CEFE, UMR 5175, CNRS - Université de Montpellier - Université Paul-Valéry Montpellier - IRD - EPHE, 1919 route de Mende, 34293 Montpellier Cedex 5, France
		\item Univ Lyon, Université Claude Bernard Lyon 1, CNRS, ENTPE, UMR 5023 LEHNA, F-69622, Villeurbanne, France
	\end{enumerate}
	{ \textbf{Running title:~} %
		Eco-evolution, timescales and modulators
	} \\ %
	\\
	{ \textbf{Author contributions:~} %
		Conceptualization: E.A.F., P.J., S.P., D.C., F.V.; Visualization: E.A.F., P.J., L.G., D.C.; Writing -- original draft: E.A.F., P.J., J.N.D., P.H., D.C., S.P.; Writing -- review \& editing: all authors.
	} \\
	\\

	\vspace{\stretch{2}}
	\begin{flushright}
		\textbf{Correspondence Details}\\
		Emanuel A. Fronhofer\\
		Institut des Sciences de l'Evolution de Montpellier, UMR5554\\
		Universit\'e de Montpellier, CC065, Place E. Bataillon, 34095 Montpellier Cedex 5, France\\
		phone: +33 (0) 4 67 14 31 82\\
		email: emanuel.fronhofer@umontpellier.fr
	\end{flushright}
\end{titlepage}

\doublespacing

\begin{abstract}
Eco-evolutionary dynamics, or eco-evolution for short, are thought to involve rapid demography (ecology) and equally rapid phenotypic changes (evolution) leading to novel, emergent system behaviours. This focus on contemporary dynamics is likely due to accumulating evidence for rapid evolution, from classical laboratory microcosms and natural populations, including the iconic Trinidadian guppies. We argue that this view is too narrow, preventing the successful integration of ecology and evolution. While maintaining that eco-evolution involves emergence, we highlight that this may also be true for slow ecology and evolution which unfold over thousands or millions of years, such as the feedbacks between riverine geomorphology and plant evolution. We thereby integrate geomorphology and biome-level feedbacks into eco-evolution, significantly extending its scope. Most importantly, we emphasize that eco-evolutionary systems need not be frozen in state-space: We identify modulators of ecological and evolutionary rates, like temperature or sensitivity to mutation, which can synchronize or desynchronize ecology and evolution. We speculate that global change may increase the occurrence of eco-evolution and emergent system behaviours which represents substantial challenges for prediction. Our perspective represents an attempt to integrate ecology and evolution across disciplines, from gene-regulatory networks to geomorphology and across timescales, from contemporary dynamics to deep time.
\end{abstract}

\newpage
\section*{Introduction}
That evolutionary and ecological change can happen on similar timescales has been known since the mid of the 20th century \citep{Pimentel1961, Chitty1967, Antonovics1976}. Interestingly, this ``old'' idea has only recently been revived thanks to conceptual advances (e.g., the genotype-phenotype map), long-term studies and advances in mathematical modelling which have made it operational \citep{Huneman2019}. Starting in the early 2000s, eco-evolutionary dynamics and feedbacks, or eco-evolution for short, have experienced an important hype \citep{Bassar2021}. Accordingly, reviews, perspectives \citep[][]{Fussmann2007, Kokko2007, Pelletier2009, Post2009, Lion2018}, special issues (BES special issue ``Eco-evolutionary dynamics across scales'' 2019) and entire books \citep{Hendry2017, McPeek2017a} have been written on the topic. However, one major question keeps coming back: What is actually an eco-evolutionary interaction? 

\citet{Hendry2017} proposes five categories: the first two are defined as eco-evolutionary ``dynamics'' where an ecological (evolutionary) change influences an evolutionary (ecological) change, but not the other way around. The third and fourth category are ``feedbacks'' which may be broad sense feedbacks where the starting and the end point of the feedback need not be identical. Fifth, the core or narrow sense eco-evolutionary feedback \textit{sensu} Hendry involves identical starting and end-points. For instance, plant seeds could exhibit morphological traits that protect them from avian seed predation. This could lead to the evolution of new beak morphologies in the birds, which may ultimately feed back on plant evolution. \citet{Hendry2017} also states that all these dynamics should be happening in ``contemporary time''. We would like to note that the word ``change'' has to be used with caution. As \citet{McShea2010} have argued, the null expectation for a biological system is continuous change rather than permanence. Variation keeps occurring, so that permanence, like the continued existence of some taxa for millions of years, is something worth explaining and may require stabilising natural selection as a cause. Therefore, in the context of eco-evolution we should understand ``change'' in its most general sense that also includes permanence, that is, zero change, as a special case.

\citet{Bassar2021} argue that the most correct and useful definition of eco-evolution is restricted to Hendry's broad and narrow sense feedbacks with an emphasis on there being ``no separation in time between ecological and evolutionary dynamics''. They also emphasize the identical ecological and evolutionary timescales implicitly assuming that both are fast. In their view, this is the only case where truly novel dynamics will emerge that cannot be explained by classical models which, as they write, usually assume a separation of timescales and weak selection, that is, small phenotypic effects of mutations \citep[][]{Lion2018}. This definition of eco-evolution has led to studies examining the conditions promoting rapid evolution, such as the genetic architecture of the traits involved \citep{Rudman2017, Yamamichi2022}. 

But what is ``rapid'' evolution? And why is ecology considered to be always fast? Timescales \textit{sensu} \citet{Slobodkin1961} can be divided into ecological, which comprises time in the 10s of generations, versus evolutionary, which lies more in the 100,000s of generations. More focused on evolutionary dynamics, \citet{Gingerich2001, Gingerich2009} differentiates between a generational timescale, which is the most fundamental timescale, and then microevolutionary and macroevolutionary timescales. Following \citet{Gingerich2001}, micro- and macroevolutionary timescales are timescales of observation and not of the actual process of evolution which happens on the generational timescale.

One of the emerging challenges is to identify how rapid evolution is relative to ecology in some pattern of interest, leading to the design of eco-evolutionary partitioning approaches \citep{Hairston2005, Collins2009, Stoks2015, Govaert2016}. Interestingly, quantification of evolutionary rates showed that these rates tend to be higher than one used to think, especially if measured on short timescales \citep[][]{Hendry1999}. All rates can be projected onto the same generation-to-generation rates if analysed correctly and evolution only seems slow on long timescales, as mentioned above \citep{Gingerich2009}. Most recently, \citet{DeLong2016} quantitatively showed, using a dataset encompassing a wide array of organisms from protozoans to humans, that evolution (rate of phenotypic change; we here remain on the level of patterns, we will discuss processes below) can be fast, but is usually slightly slower (by a factor $<$ 10) than ecological dynamics (rate of population change). These studies focus on quantitative changes of phenotypes, but qualitative changes, such as key innovations \citep{Hunter1998, Wagner2011}, might require more fundamental changes in metabolic pathways or the bodyplan, and occur less frequently, especially if historical contingencies are involved \citep[][]{Blount2008}. One may therefore again distinguish between two evolutionary timescales, more or less coinciding with the classical micro- vs. macroevolutionary differentiation.

Most of the work we have discussed so far on eco-evolution has a strong background in evolutionary biology and influences from functional or ecosystem ecology seem weak. This lack of synthesis is apparent in \citet{Hendry2017}'s book, for example, and clearly biases the existing work and hinders a full integration of ecology and evolution. While this was already noted over 10 years ago \citep[][]{Matthews2011}, limited progress seems to have been made \citep[for exceptions, see][]{Kylafis2008, Bassar2010, ElSabaawi2014, Matthews2016}. Most often, ecology, i.e., the environment, serves as a ``theatre'' for the gene-centred ``evolutionary play'' \citep{Hutchinson1965} because ecology and evolution have more or less divorced since Darwin, and the Modern Synthesis had little interest in ecology \citep{Huneman2019}. Even the ``ecological genetics'' school \citep[][]{Whitham2006} was not really interested in ecology, and no eco-evolutionary feedbacks were envisaged. Furthermore, ecology is often understood as, or reduced to, pure demography. This focus is obvious apparent in current eco-evolutionary analyses conducted by evolutionary biologists, beginning with \citet{Pimentel1961} and \citet{Chitty1967}. Conversely, it also seems that functional ecologists have had little interest in evolutionary processes \citep[][]{Loreau2010, Huneman2019}.

If we want to take a more ecosystem-level perspective, we must ask whether fluxes of matter and energy, both in trophic networks and the physical environment \citep{Corenblit2009}, may also play a role in eco-evolution. This implies that we need to understand on which timescales such fluxes happen. Using the meta-ecosystem ecology framework, \citet{Gounand2018a} have summarized available information for carbon fluxes, and show that fluxes vary widely and across orders of magnitude. Yet, often, timescales are within years, implying that ecosystem dynamics and demographic rates are comparable, and might even be intrinsically linked. Interestingly, spatial flows of matter and energy are often mediated by spatial behaviour of organisms \citep[movement, foraging, seasonal migrations, dispersal;][]{Gounand2018} which provides a mechanistic link between metacommunity and metaecosystem ecology \citep{Loreau2003, Massol2011a}. This behavioural link is especially true for taxonomically similar ecosystems (e.g., two lake ecosystems) that are more linked by dispersal than ecosystems that are biotically dissimilar (e.g., terrestrial-aquatic linkages) which may be more linked by flows of resources \citep{Gounand2018}. Of course, different organisms within an ecosystem may experience different timescales, such as bacteria versus large mammals. While interesting and potentially relevant for ecosystem dynamics this is beyond the scope of the current article.

At a more fundamental level, the abiotic environment is defined by the geological and geomorphological settings that impact abiotic ecosystem properties (e.g., via physico-chemical conditions, leaching of nutrients, fluxes and organization of mineral matter) and also impact the biotic component because geology and geomorphology provide the environmental matrix in which (meta)ecosystem dynamics play out \citep{Phillips2021}. In systems in which geology and geomorphology are the ``pacemakers'', ecological dynamics will occur at much longer timescales than demography.

\begin{figure}[h!]
	\begin{center}
		\includegraphics[width=82mm]{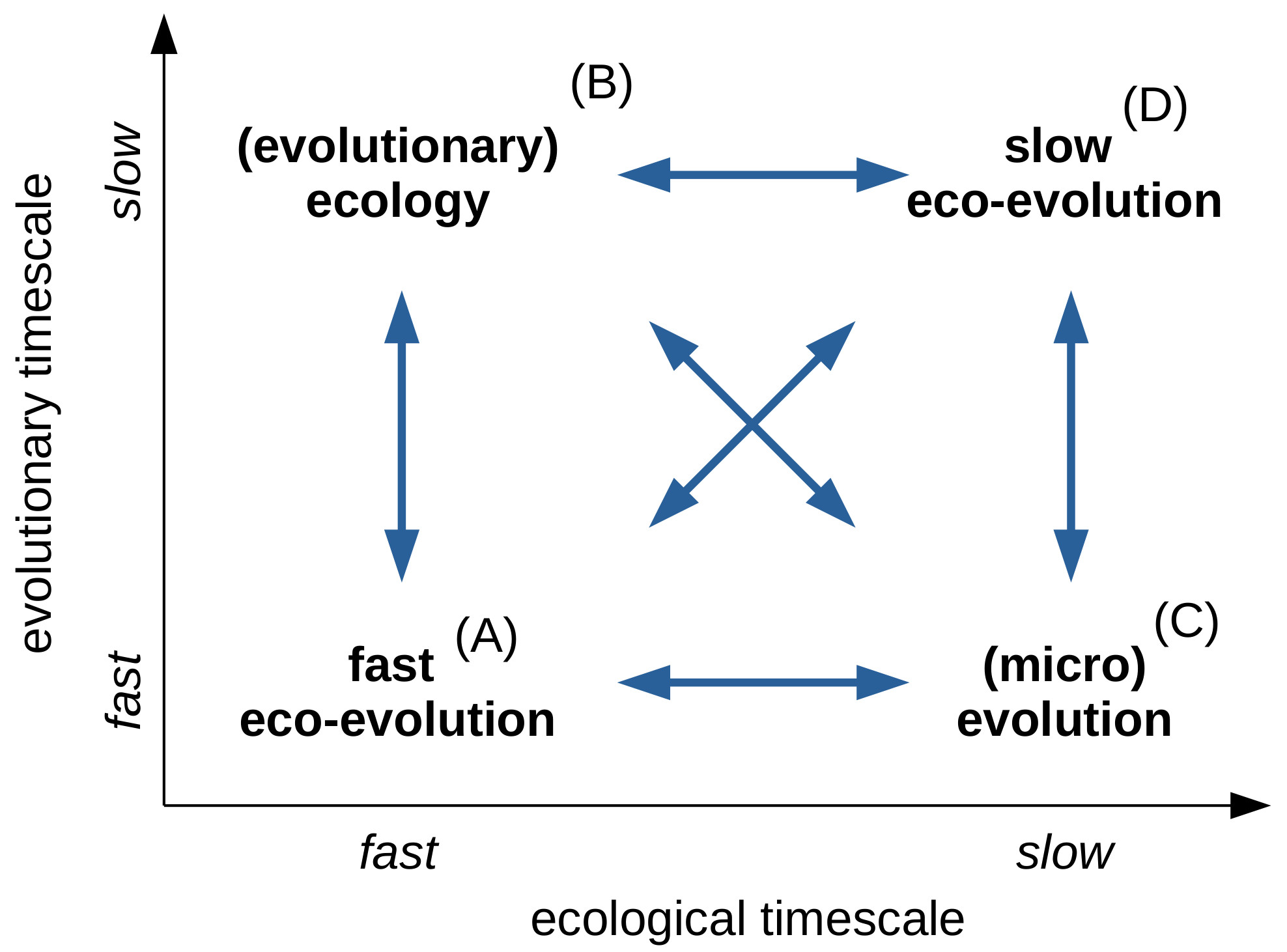}
	\end{center}
	\caption{Eco-evolutionary system states describing extremes along continuous variation of matching or mismatching ecological and evolutionary timescales. Transitions in and out of states and between states are possible via the action of modulators of ecological and evolutionary rates which slow down or speed up ecology or evolution, respectively.}
	\label{Fig:timescales}
\end{figure}

Given the central role of timescales and lack of true integration of ecology and evolution, we here propose to conceptualize eco-evolution within a two-dimensional time space. At the extremes of this time space we can identity four system states (Fig.~\ref{Fig:timescales}): A) fast eco-evolution \textit{sensu} \citet{Bassar2021} where evolution is fast enough to impact fast ecology (demography); B) classical ecology and evolutionary ecology, where evolution is too slow to immediately impact fast ecological dynamics. Note that even if the ecological dynamics results from past evolutionary changes (and vice versa), the timescales do not match; C) evolution is faster than ecology which could imply rapid adaptation to relatively slower environmental changes \citep[rapid adaptation to anthropogenic disturbance][]{Chakravarti2018, Lagerstrom2022} and, ultimately, neutral evolutionary dynamics; and finally D) both are slow such as when geomorphological conditions provide an ecological opportunity for a rare key innovation which then feeds back and impacts geomorphology \citep[][]{Corenblit2011, Butterfield2017, Pausas2022}. We recognize that these four states are artificially discrete categories, and that real situations might better be described along a continuum. 

With these four extreme system states in mind (Fig.~\ref{Fig:timescales}), we explore below eco-evolution across slow and fast timescales, bringing to the fore two aspects: 1) What is slow eco-evolution? 2) How is it possible to move between states, or, in other words, are there modulators that speed up or slow down ecological and evolutionary dynamics (or both)? We first discuss fast and slow eco-evolution, before moving to modulators of ecological and evolutionary rates and their interaction, highlighting the connection between environmental factors, demography and evolutionary processes, including the role of stochasticity. The general objective is to build an eco-evolution framework that can be used across timescales, and to derive predictions in times of environmental change.

\section*{Fast eco-evolution}
To set the scene, we first discuss one of the most classical examples of fast eco-evolution with novel dynamics emerging that has been called the ``smoking gun'' of eco-evolutionary feedbacks. This work has been synthesized by \citet{Hiltunen2014} and has been associated with eco-evolutionary interactions since the foundational work of \citet{Pimentel1961, Pimentel1968}: Cycling predator-prey dynamics, which are characterized by a quarter-phase lag between prey and predator cycles, select for (costly) defence mechanisms in the prey. As predation pressure increases with increasing numbers of predators, the prey evolves a defence mechanism, while, due to the associated costs, the undefended prey will start dominating when predator numbers are low again. This oscillation between defended and undefended prey (evolutionary change) is as fast as the demography of the predator-prey system (ecological change) which leads to an eco-evolutionary feedback in the narrow sense (eco-evolution) and a novel, emergent, system property \textit{sensu} \citet{Bassar2021}: the predator-prey system now does not oscillate with a quarter-phase lag any more but it shows anti-phase dynamics. This was first reported by \citet{Yoshida2003} for rotifers and alga, and shown to be a relatively common but overlooked feature of a lot of predator-prey time-series \citep{Hiltunen2014}, from bacteria-phage systems to insects and their parasitoids, hinting at the ubiquity of eco-evolution. Due to the similar timescales, \citet{Bassar2021} conclude that there is a very specific domain of applicability of eco-evolution: strong selection \citep[large mutational effects][]{Lion2018}, non-negligible phenotypic variances and large genetic effects on ecological variables.

Work on the Trinidadian guppy system \citep[reviewed by][]{ElSabaawi2014, Reznick2019} goes beyond a pure focus on demography and includes ecosystem-level processes in the form of nutrient recycling. This body of work shows that in low predation conditions guppy population density is higher than under high predation \citep{Reznick1982}, which leads to modifications in the ecosystem through grazing, affecting algae and invertebrates, and excretion rates, which in turn imposes selection on guppy traits \citep{Bassar2010, Bassar2012, Reznick2019}.

\section*{Slow eco-evolution}
While the above-mentioned examples of fast eco-evolution are well known, we will now discuss examples of slow eco-evolution, that is, system state D in Fig.~\ref{Fig:timescales} where both, ecological and evolutionary dynamics are slow. Stereotypically, this will imply that slow ecosystem-level processes linked to the abiotic compartment (e.g., geology and geomorphology) act as a ``pacemaker'' for ecology and that evolutionary answers to these selection pressures are slow or rare such as in key innovations that require fundamental changes in metabolic pathways or the body plan.

Going far back in time, the dynamics of oxygen on Earth provide a good example \citep{Judson2017}, especially two major oxidation events: the ``Great Oxidation Event'' slightly more than 2 Ga ago and the phase just before the Cambrian \citep{Holland2006}. The ``Great Oxidation Event'' is most likely due to cyanobacteria evolving oxygenic photosynthesis as a key innovation which raised oxygen levels in the atmosphere. This new ecological opportunity eventually led to wide-sense slow eco-evolution and permitted the emergence of eukaryotes and land plants (for a more detailed discussion see \citealt{Judson2017}; but see \citet{Mills2022} for a recent argument decoupling eukaryogenesis and oxygen levels). The rise in oxygen probably set the conditions for the explosion of animal body plans and the rise of wasteful species (i.e., species that are energetically more efficient when using resources, but are releasing more waste; \citealt{Vermeij2017}). Both aspects changed the face of the Earth since the new life forms enlarged interaction networks \citep[e.g., food chains and webs, see][]{Butterfield2007} with associated adaptations to predatory lifestyles and a major impact on the environment via bioturbation. Bioturbation is known to have driven slow eco-evolution in the oceans and on land by influencing key geomorphological and physicochemical components \citep{Johnson2002a, Meysman2006, Murray2008, Phillips2015, Butterfield2017}.

\begin{figure}[h!]
	\begin{center}
		\includegraphics[width=160mm]{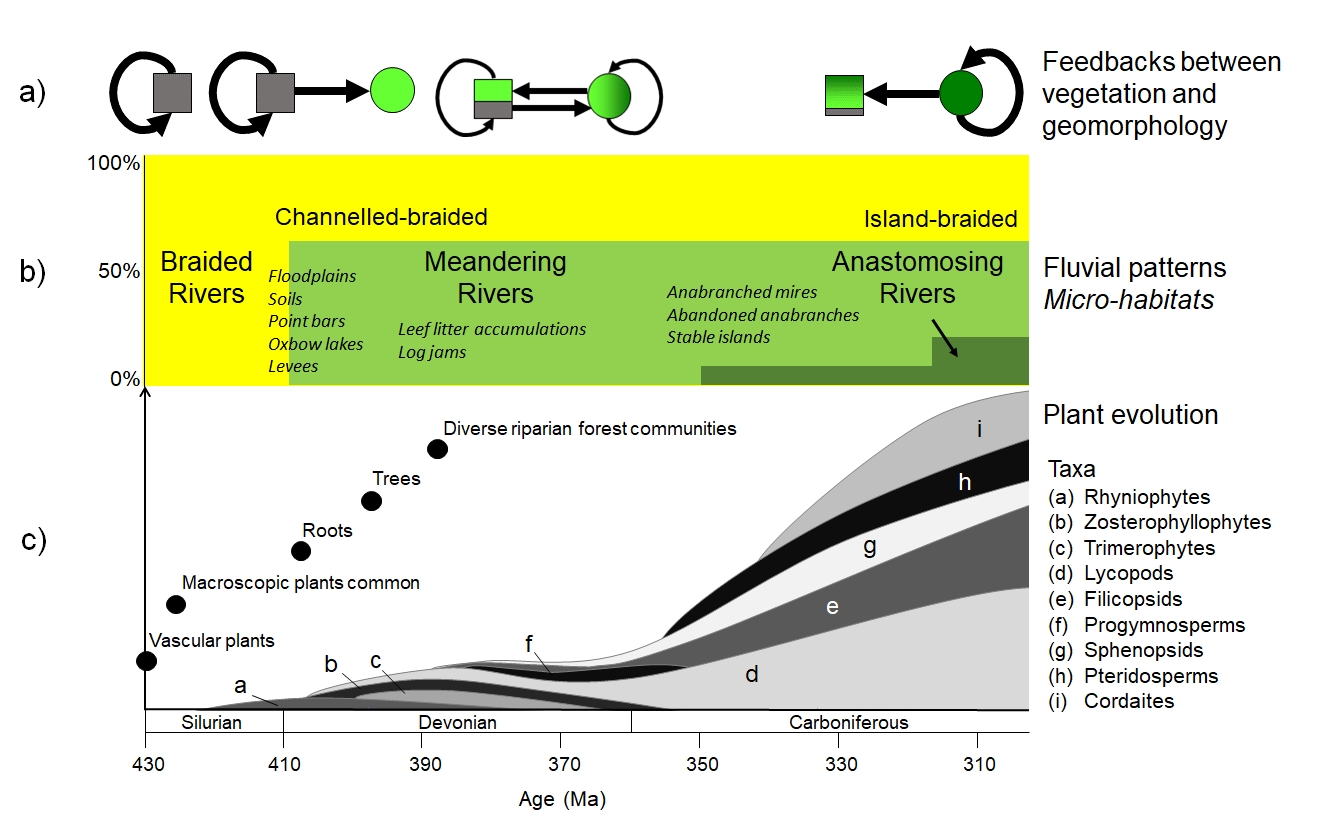}
	\end{center}
	\caption{Slow eco-evolutionary feedbacks between riparian vegetation and fluvial geomorphology during the Palaeozoic Era. From top to bottom: a) Feedbacks between geomorphology and vegetation; the squares represent geomorphology and the circles vegetation; arrow widths represent the importance of the relationships; abiotic effects are shown in grey, biotic effects in green and the dark-green colour indicates feedbacks. b) Changes in the percentage of fluvial patterns and appearance of novel micro-habitat conditions promoting selection in plant traits. c) Major aspects of plant evolution. The letters represent taxa listed in the legend (right). Adapted from \citet{Nikas1985} and \citet{Davies2013}.}
	\label{Fig:slow_ecoevo}
\end{figure}

Another well-documented example of slow eco-evolution involves riparian ecosystems which cover large areas in almost all terrestrial biomes. Plants establishing in riparian areas are subject to strong selection pressures related to flooding, drought, sediment erosion, transport and burial. At the same time, plants have an important impact on river morphodynamics via the stabilization of the substrate by their belowground organs (rhizomes, roots) and the modulation of water flow properties and sediment dynamics by their aboveground organs \citep{Gurnell2014}. As a consequence, plant traits and assemblages, on the one side, and river morphology on the other, impact each other in a feedback loop \citep{Corenblit2015}. More specifically, plant colonization of the continents began in the late Ordovician and early Silurian (444--416 Ma), with small tracheophytes colonizing coastal areas. Initially, plants evolved traits that allowed them to live in intertidal areas, followed by traits related to the new hydrodynamic, geomorphological, and physiological constraints they encountered while spreading into river corridors \citep{Bashforth2011}. The gradual evolution of riparian plant traits in the face of disturbances and stresses inherent to the coastal and fluvial environment led in turn to drastic and irreversible changes in river morphodynamics across continents throughout the Paleozoic Era \citep[Fig.~\ref{Fig:slow_ecoevo};][]{Bashforth2011, Davies2011, Gibling2014, Davies2021}. Paleozoic changes in fluvial morphodynamics led to the development of fluvial landforms that were rare or absent in the Cambrian or before, such as elevated, muddy floodplains incorporating confined sinuous channels showing at their margins steles of levees and crevasse splays \citep{Gibling2014}. It was during this pivotal period in the evolution of the biosphere that meandering and anastomosing geomorphological fluvial patterns first developed in close relation to the evolution of riparian trees with robust and deep root systems. The stabilizing and constructing effect of riparian vegetation on the floodplains fed back and caused major changes in riparian ecosystem structure and function \citep{FalconLang2011, Gibling2012}. The meandering and anastomosing fluvial patterns in particular provided new opportunities for plants and animals to evolve in a variety of patchy habitats such as main and side channels, oxbows, and to move from wet coastal to dry upland conditions, ultimately leading to the spread of seed plants and animals on hillslopes during the Carboniferous period \citep{Greb2006, Davies2013}. The fossil record indicates that the Devonian period (419 to 359 Ma ago) already encompassed numerous non-vertebrates, vertebrates, and plant communities forming complex ecosystems \citep{Labandeira1998, DiMichele2005, Kennedy2012, Wilson2020}.

Other examples involve the building of past and actual reefs, creating new geological carbonate structures in the ocean, that have subsequently served as substrate for life and the development of new ecosystems over millions of years (bioherms). Of course, reef building is a lot more complex and varied than what we can discuss here \citep[][]{Kiessling1999}.

While not being a comprehensive list, these examples highlight the existence of slow eco-evolution. The recognition that organisms affect abiotic conditions in their environment in cumulative or recurrent ways leads to many opportunities to more explicitly consider feedbacks between genes, organisms and the environment. On a slow evolutionary timescale, this may involve large-effect mutations and key innovations \citep{Hunter1998, Wagner2011} and major transitions \citep{MaynardSmith1995, Szathmary1995}, adaptive radiations and the occupation of new ecological opportunities after extinctions \citep{Stroud2016, Vermeij2017}.

All of the above mentioned examples imply that the biotic compartment impacts the abiotic compartment. This impact is often referred to as ``niche construction'', which directly connects (ecosystem) ecology and evolution. Labelled ``ecosystem engineering'' by ecologists \citep[e.g.,][]{Jones1994}, this interaction across generations changes selective pressures exerted upon the niche-constructing species and therefore induces evolution \citep{Odling-Smee2003}. More precisely, ecosystem engineering is related to the effect of a species on the structure and function of the ecosystem through modification of the physical environment without an evolutionary feedback. The latter is described by the concept of niche construction. Debates are raging about the specificity of niche construction, and its reducibility to the ordinary selective process where an allele affects the environmental parameters of the focal species \citep{Laland1999, Lehmann2007}. According to the latter view, niche construction would just be a form of extended phenotype \citep{Dawkins1982} which has led to notable debates \citep[][]{Laland2004, Dawkins2004, Laland2016}.

The concepts of ecosystem engineering and niche construction can also be found in the Earth System Sciences literature \citep[][]{Butterfield2011} since ecosystems can be seen as complex adaptive systems \citep{Levin1998, Sole2022}. A more detailed discussion of the relationship between evolutionary biology, ecology and Earth sciences is unfortunately beyond the scope of this paper.

\section*{Change of system states: what modulates ecological and evolutionary rates?}
\begin{figure}[h!]
	\begin{center}
		\includegraphics[width=160mm]{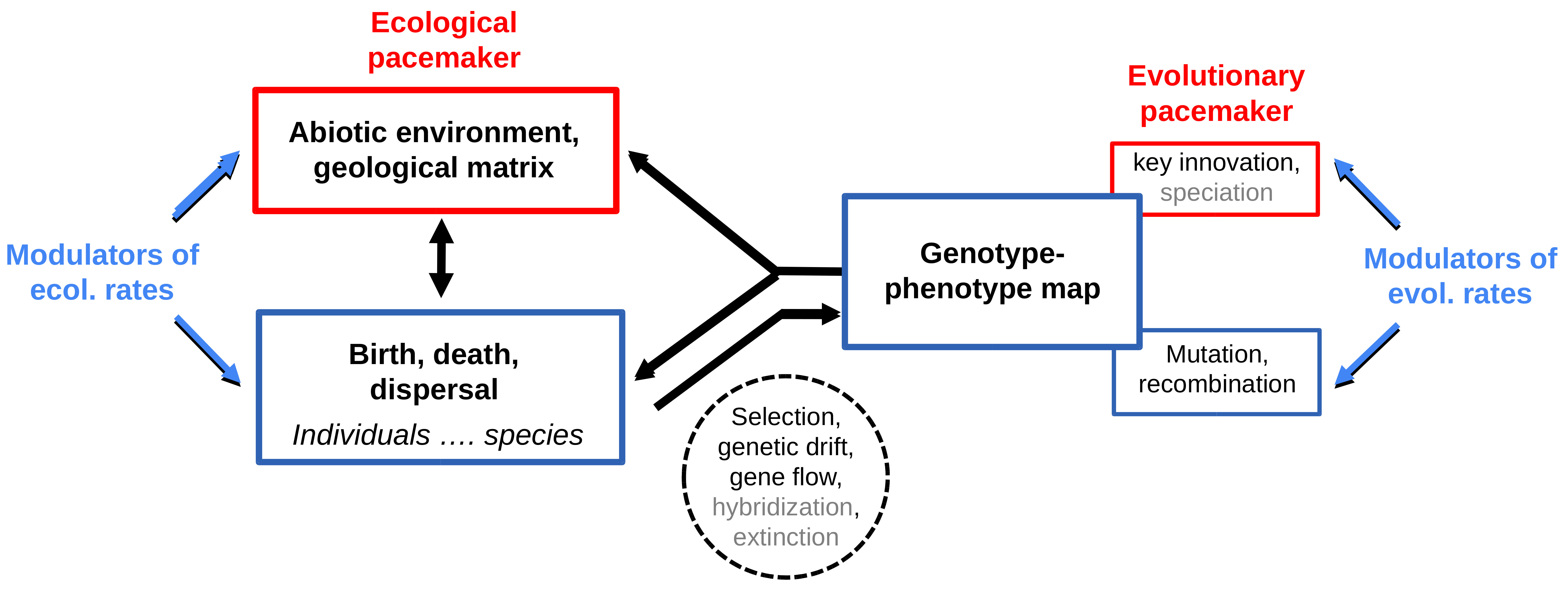}
	\end{center}
	\caption{Eco-evolutionary feedback loop with pacemakers (red) and modulators of ecological and evolutionary rates (blue). Pacemakers will define whether the fundamental dynamics are slow or fast. For instance, if slow geological processes, such as nutrient leaching, determine the pace of ecology, ecology will be overall slow. The analogous is true for evolutionary responses: if the focal evolutionary response is, for instance, linked to key innovations and speciation, this process may be overall slow. These broadly defined paces can be modulated by rate modulators. Some of these are external (e.g., temperature) or internal such as species interactions, mutation rate evolution or changes in the genotype-phenotype map. Note that the feedback loop depicted here highlights the importance of the individual level and of basic processes. We have added some emergent features such as speciation, hybridization or extinctions (grey) for clarity. Adapted from \citet{Govaert2019}.}
	\label{Fig:ecoevo_modulators}
\end{figure}

As mentioned earlier, the four system states identified in Fig.~\ref{Fig:timescales} are extremes along a continuum of matching or mismatching rates of ecological and evolutionary processes. In reality, many situations may exist in between. We might even expect transitions between these states (Fig.~\ref{Fig:timescales}) or at least movement along the axes when ecological and evolutionary rates change. For example, stressful environmental factors may decrease birth rates and thereby slow down ecological dynamics. By analogy, mutation rates may be impacted, for example, by mutagens, and speed up evolutionary dynamics in response to a given selection pressure. We may even imagine transitions from the slow to the fast eco-evolution regime. In the following, we will refer to the factors responsible for changes in the ecological or evolutionary rates as ``rate modulators'' (Fig.~\ref{Fig:ecoevo_modulators}). While certain processes may modulate both ecological and evolutionary rates, we first focus on distinguishing between modulators of ecological rates and modulators of evolutionary rates (Fig.~\ref{Fig:ecoevo_modulators}) and will discuss interactions later. We here do not provide an exhaustive list of all potential rate modulators, but rather want to illustrate the role of rate modulation for eco-evolution.

\subsection*{Modulators of ecological rates: changing the speed of ecology}
Overall, ecological rates will either be determined by demographic rates (rates of birth, death, and dispersal) or by abiotic components (current environmental conditions but also physico-chemical conditions, leaching of nutrients, fluxes and organization of mineral matter) as discussed above (Fig.~\ref{Fig:ecoevo_modulators}).

Ecological rate modulators are well-known, and involve classical stressors such as temperature, pH, salinity, precipitation, and other environmental factors. These will be relevant for fast eco-evolution, but oxygen levels in the atmosphere, for example, are also an ecological rate modulator which have impacted the evolution of biodiversity in deep time (see above). Interestingly, oxygen is a rate modulator that is produced and consumed by organisms, which may hint at complex feedbacks. The effect of these stressors may increase after natural catastrophic events (e.g., asteroid impacts, volcanic activity), changes caused by human activities (e.g., increases in pH or temperature) or due to slower tectonic or astronomic forcing (e.g., orogenesis or continental fragmentation and Milankovitch cycles at the basis of long-term climatic variations that deeply impact biomes and their distribution; also long-term volcanic activity, e.g., the formation of Deccan Traps and their potential role in the Cretaceous -- Paleogene extinction event). Importantly, the form of their effects can vary broadly, from linear, monotonic to unimodal or u-shaped. If effects are extremely non-linear, these have usually been described as tipping points implying alternative stable states \citep{Scheffer2001, Drake2020}. We start with discussing what modulates demographic rates and then focus on modulation via the abiotic environment (Fig.~\ref{Fig:ecoevo_modulators}). As ecology, if defined by demographic rates, may generally be seen as (slightly) faster than evolution, any ecological rate modulator that slows down ecological dynamics may lead to an increased likelihood of eco-evolution and, therefore, emergent system properties.

\subsubsection*{Modulation of birth and death rates}
Temperature is a well known ecological rate modulator as it regulates the metabolism of organisms, and hence can alter birth and death rates \citep{Brown2004, Gillooly2001}. Biological rates may scale exponentially within a certain range of temperature, but will globally scale with a unimodal relationship. If temperature alters birth and death rates, it can alter the ecological dynamics governed by these processes and consequently alter the rate of demographic processes. This is highly relevant in the current context of ongoing global changes, in which temperatures are expected to rise quickly and globally \citep{PerkinsKirkpatrick2020, MassonDelmotte2021}, and may thus push systems into alternative eco-evolutionary states by changing ecological rates. Global change-induced rises in temperature may speed up or slow down ecological processes, depending on whether environmental conditions are above or below the focal organism's thermal optimum. For example, in consumer-resource systems, both foraging rate and consumer-resource interaction strength depends on temperature \citep{Dell2014a, Gilbert2014, Hamann2020, Synodinos2021}. If higher temperatures speed up birth rates of prey or predator species, this could lead to fast eco-evolution or, on the contrary, disrupt fast eco-evolution, depending on the rate of evolution. More generally, if a system already has matching ecological and evolutionary rates, the consequence will be a reduction of the potential for eco-evolution because ecological timescales will become faster or slower than evolutionary ones.

Similar to temperature, other physicochemical stressors (e.g., pH, salinity, humidity, pollution) that have linear or non-linear relationships with demographic rates may shift eco-evolutionary system states. Predicting whether such stressors act as positive (speeding up) or negative (slowing down) rate modulators will depend on the shape of the response curve, but, if all stressors ultimately have a negative impact on population growth, the consequences in terms of matching or mismatching eco-evolutionary timescales will depend on whether the stressor acts via decreasing birth rates (slowing down ecology) or increasing death rates (speeding up ecology; see also \citealt{Boyce2006}, \citealt{Lawson2015}).

Stressors may of course exist outside of the global change context. Especially gradients of stress, such as large scale, latitudinal or altitudinal temperature gradients may be responsible for determining species ranges exactly via ecological rate modulation. For instance, the ``Species-Interactions Abiotic-Stress Hypothesis'' (SIASH, reviewed by \citealt{Louthan2015}) states that stressful range edges are defined by abiotic forces and that non-stressful edges are more defined by species interactions (for a focus on facilitative interactions see the ``Stress Gradient Hypothesis'' by \citealt{Bertness1994} which has somewhat opposing predictions). One potential mechanism for the SIASH is the reduction of inter-individual interactions at stressful margins via reduced densities, which leads to a spatial gradient in ecological rates. As a consequence, such large-scale gradients may lead to geographical hot- and cold-spots of eco-evolution. Patterns can of course be complexified due to dispersal, for instance.

Beyond the abiotic environment, biotic interactions themselves may act as rate modulators (Fig.~\ref{Fig:ecoevo_modulators}). For example, the presence of predators may induce variation in morphological and life-history traits, as shown by studies on the effect of crayfish on maturation age in freshwater snails \citep[][]{Hoverman2005, Covich2010}. More generally, predator defence mechanisms or competitive ability may be linked to demographic rates directly (via density regulation in the case of competition; see e.g., \citealt{Fronhofersubmitted}, \citealt{Siepielski2020}) or indirectly (via costs of anti-predator mechanisms, for example; \citealt{Urban2007}) and modulate eco-evolutionary rates.

\subsubsection*{Modulation of dispersal}
External factors may also impact the third demographic rate: dispersal. Fragmentation, for instance, and loss of habitat in general, may reduce effective dispersal rates and thereby reduce the speed of ecological dynamics \citep{Legrand2017}. Conversely, the rewiring of dispersal networks \citep{Bullock2018} may facilitate dispersal and speed up ecological dynamics. Dispersal may also be modulated via biotic interactions. Theory indicates that dispersal should globally increase with intra- and interspecific competition \citep{Metz2001, Poethke2002}. Empirical evidence indicates that, in a food-web context, both bottom-up and top-down effects will determine dispersal rates \citep{Fronhofer2018, Coteaccepted}.

\subsubsection*{Modulation via the abiotic environment}
In the above sections we have highlighted processes that may directly modulate demographic rates which may be most relevant when ecological dynamics are globally fast. For modulation of globally slow ecological processes we propose to differentiate three broad categories: 1) Slow ecological processes with external slow forcing, e.g., glaciations over the last three million years. Here, forcing was external via the position of the Earth and temperature, and other environmental parameters, changed slowly allowing organisms to potentially track climatic zones geographically and occupied ``refuges''. The slowness of these processes likely prevented eco-evolution, while local adaptation to refuges was possible. Another example is the colonization of land ca. 500 Ma ago where the abiotic environment was defined by available space, which led to ecological opportunities and subsequent major key innovations \citep{Vermeij2017} leading to slow eco-evolution. 2) Slow processes with internal (biological) slow forcing which includes the great oxidation events discussed earlier. The rise in oxygen was due to internal forcing with bacterial activity slowly rising and coinciding with the rate of key innovations which implies slow eco-evolution. 3) Finally, slow processes following catastrophic events, such as events that happened at the Cretaceous -- Paleogene boundary and the massive extinction of non-avian dinosaurs and many other taxonomic groups, leaving room for birds and mammals \citep{Longrich2012}. Similarly to the internal forcing example, here, ecological opportunities arise paving the way for key innovations and further slow eco-evolution.

\subsection*{Modulators of evolutionary rates: changing the speed of evolution}
Evolutionary changes are determined by four processes: selection, genetic drift, gene flow (including across species through hybridization and subsequent introgression) and mutation (including recombination). At the species-level one may also mention extinctions and speciation. While we do not intend to wade into a discussion on the link between processes that are classically understood as microevolutionary and processes labelled as macroevolutionary, we here assume that species-level patterns, such as speciation are driven by underlying processes such as (the lack of) gene flow, for example.

All of these processes clearly influence the rate of evolution. However, a discussion of evolutionary rate modulators is complicated by the fact that such changes in evolutionary rates are also mediated by changes in ecological rates, particularly, in the case of selection, gene flow and genetic drift (Fig.~\ref{Fig:ecoevo_modulators}). This is because selection pressures, gene flow and drift are the evolutionary forces that are most likely directly defined by ecology (Fig.~\ref{Fig:ecoevo_modulators}). Note that this is not completely true as discussed, for example by \citet{Futuyma2010} who points out that selection may also be ``internal'' via developmental effects. In Fig.~\ref{Fig:ecoevo_modulators} this may be captured by the genotype-phenotype map. Gene flow is further complicated by the fact that it is mediated by dispersal which is an ecological rate but at the same time a trait with a genetic basis \citep{Saastamoinen2018} which allows for complex, second-order evolutionary processes which we will discuss below.

In general, the rate of an evolutionary response is contingent on the amount of heritable  phenotypic variation that aligns with the optimum that is selected for, potentially determined by ecological change \citep{Schluter1996} and the ability of a biological system to show an adaptive response. We will here focus on modulation of evolutionary rates in the context of evolvability, and its evolution \citep{Payne2019}. We do not discuss epigenetic variation in any detail but some of the reasoning below may also apply. 

The term ``evolvability'' has been used differently depending on timescale and context \citep{Pigliucci2008a, Riederer2022}. On short timescales, evolvability is defined as standing genetic variation \citep{Houle1992}. If such variation is already present, then a rapid evolutionary response is expected in response to rapidly changing ecological conditions, that is, fast eco-evolution. A well-documented example is the Atlantic killifish which was shown to have adapted extremely rapidly (and repeatedly) to high levels of industrial pollutants in estuaries of the north-eastern coast of America \citep{Reid2016, Lee2017}.

At intermediate timescales, mechanisms that generate standing genetic variation, termed variability (potential to vary; \citealt{Wagner1996}), contribute to evolvability. The potential to generate variation depends not only on mutation rates but also how these mutation rates impact fitness \citep{Riederer2022}. Thus, mechanisms that generate, deplete or maintain variation are critical to understanding how evolutionary rates are modulated. We will discuss modulators of evolutionary rates in two sections, first focusing on external modulators, such as temperature or stressors and, second, highlighting internal modulators, for instance, related to genetic architecture.

\subsubsection*{External modulation}
External modulation of evolutionary rates may be due to processes very similar to those discussed above for ecological modulators. Focusing on variability, temperature, for instance, has been shown to impact spontaneous mutation rates in a u-shaped manner, with increased mutation rates at both low and high temperatures \citep{Waldvogel2021}. Other stressors, such as UV or chemical contaminants, are also well known to impact mutation rates \citep[see e.g.,][]{LopezBarea1998, Somers2002, Bickham2011, Saaristo2018}.

In analogy to ecological rate modulation and focusing on variation, influx of novel genes or genotypes via gene flow will be modulated negatively by fragmentation and positively by rewiring of dispersal networks \citep{Bullock2018}. Gene flow may not only bring new alleles, but also lead to new allelic combinations through hybridization and possible introgression. However, the question is whether such inflow will increase or decrease the rate of evolution. Isolation may indeed promote local adaptation, while high gene flow may reduce local adaptation and could slow down evolutionary responses \citep{Massol2011, Laroche2016}.

Finally, in the context of slow eco-evolution, ecological opportunities, which may often be provided by previous (mass) extinction events, can promote key innovations, diversification and speciation. Therefore, such extinctions events may also be seen as external modulators of evolutionary rates. Large-scale latitudinal gradients of speciation rates as reported by \citet{Weir2007} or \citet{Rabosky2018}, for instance, also provide examples of evolutionary rate modulation, potentially driven by associated major environmental factors.

\subsubsection*{Internal modulation}
Modulation of evolutionary rates may however be a lot more complex. Specifically, evolution of mutation rates, as well as changes in the structure of the genotype-phenotype map, can change evolutionary rates. 

Mutation rates are subject to evolution with, for example, selection favouring overall low mutation rates in stable environments for well adapted populations because the majority of mutations have harmful effects \citep[see also][for hypermutator fates in \textit{E. coli}]{Wielgoss2012}. Evolution of mutation rates might be regulated by the physiological cost of maintaining mutations at low level or by genetic drift \citep{Sniegowski2000, Lynch2016}. Evolution of mutation rates is probably rather to be expected on longer timescales or in alternating bouts of high and low mutation rates \citep{Giraud2001}. This is not the case for recombination and hybridization which may modulate rates on both short and long timescales impacting adaptation rates but also potentially leading to fast innovation. Of course, the effect of mutations will depend on their effect size, with large effect mutations having the potential to lead to innovations. 

Recombination rate may evolve under stressful environmental conditions, most notably in species with flexible sexuality including transition from asexual to sexual reproduction \citep{Burke2017,Gerber2018, Moerman2020a}. Of course, asexuality can also be triggered by demographic conditions such as low density, for example, and not because of selection. In addition, hybridization processes (i.e., interspecific gene flow) can speed up evolutionary rates, which is illustrated by the Gulf killifish. This species rapidly adapted to the extremely polluted environment of the Houston harbour thanks to an adaptive introgression by the Atlantic killifish. This introgression occurred because of secondary contact between the two killifish species most probably due to human-assisted transport \citep{Oziolor2019}. This latter example also shows that evolutionary rates can be modulated by ecological rate modulators, here human-driven dispersal.

Ultimately, the fitness impact of mutations depends on the structure of the genotype-to-phenotype map which includes mutation effects, pleiotropy and epistasis. The structure and properties of this map \citep{Nichol2019} are thus relevant to defining evolutionary rates at varying timescales. Since in realistic maps multiple genotypes can correspond to one phenotype, robustness to mutation can emerge. While the potential to generate more variation might lead to rapid evolutionary responses at intermediate timescales, on longer timescales, robustness to mutation (genetic canalization) can lead to large leaps in evolution. This is because mutationally robust genotypes can neutrally explore genotypic space without any changes in the phenotype \citep[see][]{Ciliberti2007}. On longer timescales, modularity in developmental systems promotes innovations \citep{Wagner1996}. Hence robustness on short timescales allows for evolvability in longer timescales \citep{Wagner2011}. Evolvability, in turn, allows species to persist in the midst of ecological change. Thus, when taken into account in evolutionary modelling, the genotype-phenotype map explains a form of short-timescale robustness, which, in turn, stabilises ecological change at larger scales due to the evolvability of the species. This mutational robustness acts as an evolutionary rate modulator. Under conditions of rapid ecological change, a loss of robustness (decanalization) can speed up evolutionary responses \citep{Deshpande2022}. Further, phenotypic plasticity can also modulate evolutionary rates (genetic assimilation or the accumulation of cryptic genetic variation; e.g., \citealt{VanGestel2016}).

A majority of the above-mentioned evolutionary rate modulators allow to speed up the pace of evolution. Since evolution seems to be slightly slower than ecology \citep{DeLong2016}, these processes may bias dynamics towards the fast eco-evolution \textit{sensu} \citet{Bassar2021}.

\subsection*{Interactions between ecological and evolutionary rate modulators}
The action of rate modulators is likely a lot more complex than what we have described above, in particular because they may simultaneously impact ecological and evolutionary change (Fig.~\ref{Fig:ecoevo_modulators}).

\begin{figure}[h!]
	\begin{center}
		\includegraphics[width=82mm]{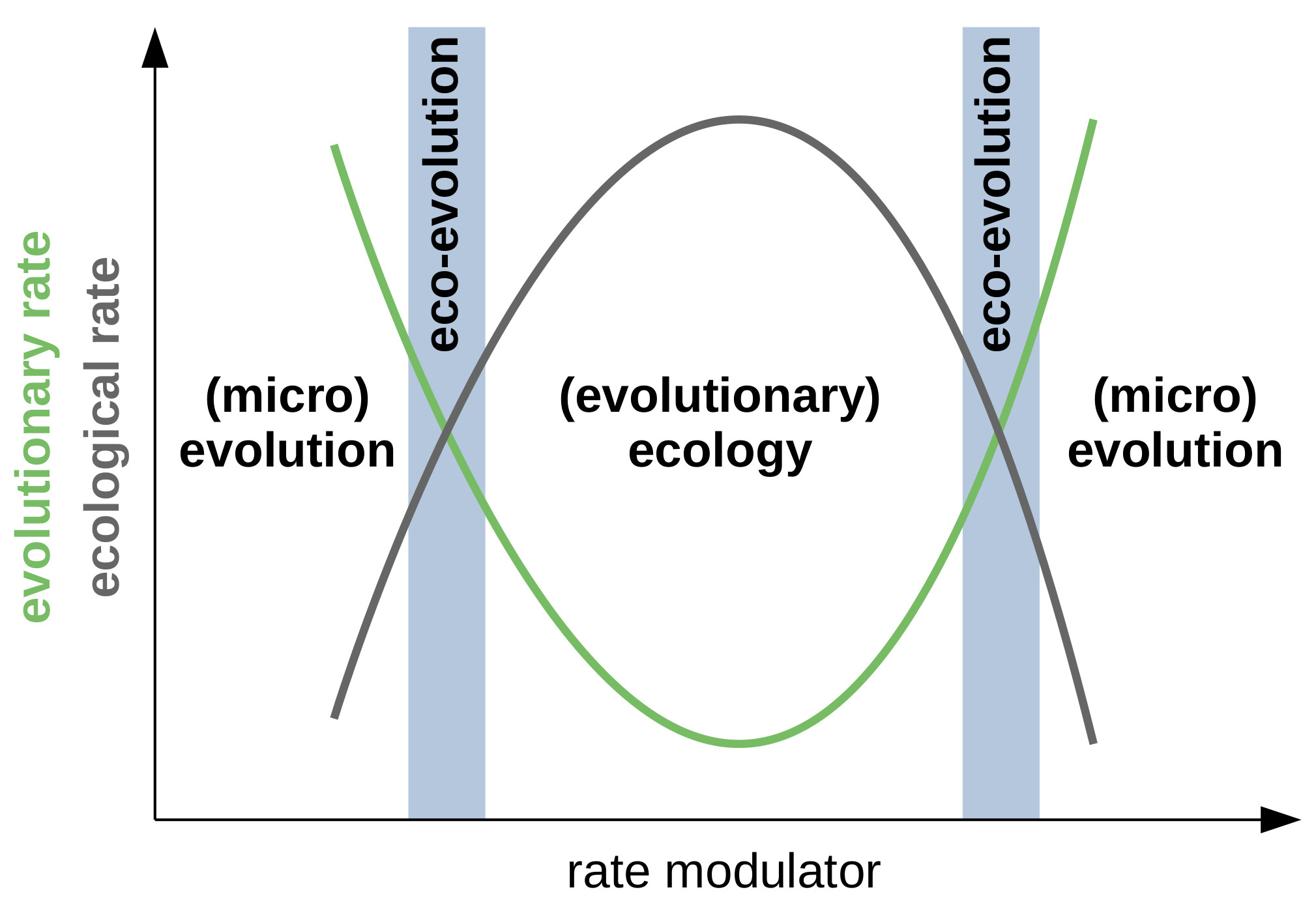}
	\end{center}
	\caption{Possible interactions between modulators of ecological and evolutionary rates leading to changes in system states from evolution to eco-evolution via ecology back to eco-evolution and evolution (from left to right). Concretely, the rate modulator here is temperature which increases, for example, due to anthropogenic activities from left to right (alternatively, this could also be a latitudinal gradient). The rate modulator has a concave (convex) effect on the ecological (evolutionary) rates. For temperature, the grey (ecological) curve represents a thermal performance curve and temperature effects on population growth rates, while the green curve represents the impact of temperature on mutation rates, for example. Crossing lines indicate similar rates which implies that emergent, eco-evolutionary phenomena become possible. Of course, rate-modulator relationships can take different forms which will determine in which state the ecological system is (see Fig.~\ref{Fig:timescales}). Note that the example mentioned here is overall in the fast eco-evolution realm, but analogous patterns in the slow eco-evolution case are also possible.}
	\label{Fig:modulator_interaction}
\end{figure}

One example is again temperature which we have discussed in both ecological and evolutionary contexts. Interestingly, it will modulate ecological rates globally following a concave relationship (classical hump-shaped thermal performance curve) while the evolutionary modulation of mutation rates may be convex (u-shaped; \citealt{Waldvogel2021}; for recombination see \citealt{Morgan2017}) as discussed above (see Fig.~\ref{Fig:modulator_interaction}). Considering that evolutionary rates are on average smaller than ecological ones \citep{DeLong2016} this leads to a transition from slow ecology with fast evolution at low and high temperatures to fast ecology and slow evolution at intermediate temperatures, close to the optimal temperature. Strong eco-evolution can therefore be predicted to occur at the intersection of these curves, namely to the left and the right of the mode of the thermal performance curve (Fig.~\ref{Fig:modulator_interaction}). In a climate change context, with extreme warming and temperatures beyond an organism's optimum, we can therefore predict the increased occurrence of eco-evolution.

More generally, ecological rate modulators will also impact evolution, depending on whether they primarily act on birth or on death rates. If stressors mainly act via a decrease in birth rates \citep[see e.g.,][]{Aulsebrook2020} evolutionary rates will also be decreased because less births imply less input of novel mutations. By contrast, an effect on death rates \citep[e.g.,][]{Pardo2017} may have no impact or increase the strength of selection. Dynamics may get even more complex if one thinks about processes like senescence, that will increase selection early on in the life-cycle \citep{Rose1991}, or trade-offs and trait correlation such as the competition-colonization trade-off \citep{Cadotte2006, Cadotte2007}, for example. The effect of ecological rate modulators on evolution may also be more indirect. It has for instance been shown that mutation rates can depend on population density \citep{Krasovec2017}. Therefore, any ecological rate modulator that changes emergent population densities also immediately has the potential to impact mutation rates.

As mentioned above, regardless of what demographic rate is impacted by the modulators, selection pressures will be changed, thereby indirectly linking ecology and evolution. One example is the variability of selection pressures due to frequency-dependent selection. In this case, an ecological rate modulator may imply a feedback on evolutionary change that decreases the intensity of selection, which in turn impacts ecological change, slowing it down. This eventually lets evolutionary change occur in a more constant ecological setting.

At the intersection between ecological and evolutionary rate modulators, dispersal and its drivers may become especially prominent. Clearly dispersal can be a key player in eco-evolution \citep{Govaert2019}: it is itself an ecological rate, it defines gene flow and therefore the pace of evolution and, finally, it has a genetic basis \citep{Saastamoinen2018} and can itself be subject to evolution \citep{Bowler2005, Ronce2007}. This combination of roles may lead to complex eco-evolutionary feedbacks \citep[for a discussion of temperature and fragmentation effects on eco-evolution, see][]{Faillace2021}.

Last, we would like to mention that all rates and modulators mentioned above may be context-dependent, such as dependent on body sizes or complexities of the ecosystem and therefore exhibit plasticity. While certainly interesting, these complexities are beyond the scope of this work. In conclusion, changes in ecological and evolutionary rates are highly coupled, and causation is not always easy to disentangle, calling for detailed analyses of the mechanisms involved in eco-evolution and their respective strength (Fig.~\ref{Fig:ecoevo_modulators}, \ref{Fig:modulator_interaction}).

\subsection*{Introducing stochasticity}
Up to now, we have not, at least not explicitly, considered stochasticity, although it plays a key role in both evolution \citep{Lenormand2009}, hotly discussed in the wake of \citet{Gould1989}'s ``replaying the tape of life'' idea, and ecology \citep{Shoemaker2020}, and presumably also for eco-evolution. Stochasticity is a hallmark of living systems, it can be demographic or environmental, and we outline some of its impacts below.

Environmental stochasticity is perhaps, in our context, the most straightforward to consider since it affects both the evolutionary and ecological dynamics. Environments are often spatially heterogeneous, and may fluctuate in time in a more or less predictable fashion. Of importance here is that the abiotic side of environmental stochasticity may affect living entities, from genes to ecosystems, at various timescales on the fast to slow continuum of eco-evolution. For example, weather variation or the occurrence of floods, storms and fires can be sources of stochasticity at short timescales. Of course, mitigating environmental stochasticity is part of the selective process at these timescales \citep{Lenormand2009}, for example through bet-hedging strategies. A less common source of stochasticity comes from rare large-scale events, such as volcanic eruptions with global effects or asteroid impacts \citep{Hoffman1998}, leading to extreme consequences on living beings, such as mass extinctions \citep{Raup1982, Longrich2012}, and no less extreme evolutionary consequences such as radiations \citep{Penny2004}.

Ecological rate modulators and changing demographic rates will also directly impact demographic stochasticity. Even if the equilibrium density is not impacted, higher underlying birth and death rates will increase demographic stochasticity due to increased turnover at population equilibrium. Even more extreme, increasing demographic rates are known in discrete-time systems to lead to deterministic chaos \citep[][for an experimental demonstration of chaos see \citealt{Becks2005}]{Hassell1975, Hassell1976}.

Besides these direct impacts on variance in population dynamics, ecological rate modulators can also have indirect impacts on demographic stochasticity, via effects on (equilibrium) population sizes. For instance, equilibrium population sizes have been shown to decrease with increasing temperature \citep{Bernhardt2018}, which implies that demographic stochasticity will become relatively more important with increasing temperature, for example.

Beyond obvious consequences for stability, risk of extinction and predictability of systems, these ecological effects also impact evolution via increased genetic drift, if population sizes are small enough. We know that drift may play a (non-directed) positive role in evolution \citep{Lenormand2009}, for example by purging deleterious mutations under specific conditions \citep{Glemin2003} or by impacting mutation rates \citep{Lynch2016}. Over much longer timescales, drift may shape genome evolution with differential effects depending on organism size \citep[effective size;][]{Lynch2007}. Similarly, genetic diversity in animals seems to be strongly linked to a slow--fast continuum of life-history strategies, such that demography and drift might also be at play here \citep{Romiguier2014}. Clearly, at larger scales, founder effects may play important roles in (island) biogeography, for example. Especially when coupled with rare, long-distance dispersal such events can shape biogeographic patterns \citep{Gillespie2012}.

Genetic drift may also be relevant in a spatial context via its spatial analogue, gene surfing. Gene surfing is a process by which neutral or even deleterious alleles can increase in frequency at expanding range fronts thanks to sequential founder events \citep[for a review see][]{Miller2020}.

\section*{Discussion and perspectives}
The field of eco-evolutionary dynamics has, in recent years, suffered from a lack of clear definition to the point where every study at the interface between ecology and evolution has used the label eco-evolutionary dynamics or feedbacks. \citet{Bassar2021} have addressed this problem recently and have clearly argued under what conditions eco-evolution yields different predictions and insights in comparison to more classical work in evolutionary ecology. These authors focus on the emergence of novel patterns that would not be within the realm of possible explanations in either purely ecological or purely evolutionary contexts. Note that the term ``emergence'' is often used rather loosely implying only the appearance of a pattern of some kind. Strictly speaking, emergence refers to  phenomena or processes that are underivable from their components, or unpredictable from the laws that govern basic phenomena \citep[see][]{Bedau2008, Huneman2008}. \citet{Grantham2007}, for example, showed that some biogeographical dynamics are indeed emerging from regional ecological processes.

Here, we revisit the definition of eco-evolution and argue that, for emergence to happen, similar timescales are enough, and do not necessarily require fast ecology and evolution (Fig.~\ref{Fig:timescales}). We show that eco-evolution can very well embrace the ecosystem level as well as geology and geomorphology while keeping its focus on emergence. Indeed, slow eco-evolution allows us to understand emergent phenomena over longer timescales, such as the trajectories of ecosystems in which abiotic components are coupled to the biological activities of ecosystem engineers. In these systems, adaptive changes of organisms to the environment participate in the formation of emerging patterns and ecosystem self-organization.

Of course, reciprocal feedbacks can also occur across different timescales \citep[e.g.,][Fig.~\ref{Fig:timescales}]{Lion2018, Govaert2019} and interesting phenomena can be studied in that context. We here do not argue that these lines of research are not valuable, on the contrary, we would like to call for a broader integration of ecosystem ecology, geology, paleontology and evolutionary biology. Such an integrative approach can help, not only understanding the past and current dynamics of biodiversity, but also to tackle the challenges of the future associated with massive urbanization and global change. As described above, emergent phenomena are by definition underivable from their components which represents important challenges for theory and modelling and for moving from a descriptive to a more predictive science of the environment \citep{Mouquet2015, Urban2016, Houlahan2016, Yates2018}. 

Besides expanding the eco-evolution framework across timescales, we highlight a central point, that is especially relevant in changing environments: biological systems do not need to be associated with a specific timescale forever. Rates can be modulated (Fig.~\ref{Fig:ecoevo_modulators}), that is, sped up or slowed down by what we term ``rate modulators'' above, which move biological systems along the ecological and/or evolutionary timescale axes. While we could not provide a comprehensive list of such modulators, the main point we would like to underline is the dynamical nature of eco-evolutionary systems.

Central questions that our work implies are: Is eco-evolution likely, is it an (evolutionary) repellor or maybe an attractor? Where are real biological systems in the diagram of states represented in Fig.~\ref{Fig:timescales}? While we can only speculate about the answers to these questions, we would like to highlight a few important takeaways. Global change, such as increased temperature, landscape fragmentation or sea-level rise, has the potential to impact biological systems and move them in or out of their current eco-evolution state. Specifically, temperature increases above an optimum have the potential to slow down ecology and speed up evolution forming eco-evolutionary feedbacks and, therefore, making emergent phenomena more likely. At the same time, many other environmental changes are becoming stronger (e.g., landuse change, urbanisation, pollution). While many ecologists may assume that these pressures are occurring too fast and are too strong for evolution to be relevant, these strong environmental changes actually can exert strong selection pressures on organisms, prompting fast evolutionary responses \citep[see][for a discussion of eco-evolution and coral conservation]{Colton2022}. Together with evolutionary rate modulators, such as the capacity of genetic architecture to speed up evolutionary responses via decanalisation, strong environmental pressures may be exactly creating the context that moves biological systems into the eco-evolution realm. In combination with decreasing population sizes and increasing stochasticity, this represents an important challenge for predicting the future of ecosystems. On a more speculative note, slow eco-evolution may be relevant when trying to envision changes over the very long term, such as the future of the Earth including (or not) humankind after a sixth mass extinction and potential novel adaptations including radiations of new biological groups.

With the increasing availability of large amounts of data and immense computing power, machine learning and artificial intelligence have been advocated as a solution for improving predictability in ecology and evolution \citep{Peters2014, Rammer2019}. However, the increased occurrence of emergent phenomena under eco-evolution implies that these correlative approaches may reach their limits because of their lack of a mechanistic underpinning. It is currently unclear what kind of models are needed in such non-analog, that is, new, conditions \citep{Yates2018}. \citet{Urban2016} argue that mechanistic models including species interactions, dispersal, demography, physiology, environment and evolution are promising. Network-based approaches can represent a productive way forward in this context. As \citet{Melian2018} highlight, multilayer networks \citep{Pilosof2017} can be used to capture eco-evolution from gene-networks to networks of ecosystems. While such approaches are regularly used in ecology they are less often found at the interface between ecology and evolution \citep[but see e.g.,][]{Deshpande2022}.

\subsection*{Conclusions}
In conclusion, we have discussed that, while keeping the definition of eco-evolution by \citet{Bassar2021} based on emergence, one can integrate ecology and evolution more broadly when focusing on the entire spectrum of timescales of both ecology and evolution. Our perspective therefore highlights important possibilities for a better integration of ecosystem sciences, geology, geomorphology and evolutionary biology.

Our timescale-based perspective further shows that biological systems need not be frozen in state space. Rather, rate modulators may move them into and out of slow or fast eco-evolution regimes. Especially in the context of global changes, being aware of the fact that emergent eco-evolutionary behaviours may arise, or be prevented, is crucial for developing a more robust and predictive eco-evolutionary science. 

Our work highlights two main challenges: 1) understanding current and past eco-evolution requires bridging across all disciplines studying ecology and evolution, regardless of whether they focus on the presents or on the past, which 2) leads to an important challenge for prediction. As a consequence, future eco-evolutionary work has to be more integrative, including levels of complexity from gene networks to networks of ecosystems. The speculative nature of many ideas discussed above clearly calls for more formal theory and experimentation in order to make concepts operational.

\section*{Acknowledgements}
We thank Pierre-Olivier Antoine and Olivia Judson for valuable comments on a previous manuscript version. P.J., E.A.F, S.P., D.C. and F.V. are supported by the CNRS Institute for Ecology and Environment. This work was supported by a grant from the Agence Nationale de la Recherche (No.: ANR-19-CE02-0015) to EAF. This is publication ISEM-YYYY-XXX of the Institut des Sciences de l'Evolution --- Montpellier.

\section*{Conflict of interest}
The authors declare no conflicts of interest.

\newcounter{MyBibCount}\providebool{MyRefNumbers}


\begin{thebibliography}{186}
	\expandafter\ifx\csname natexlab\endcsname\relax\def\natexlab#1{#1}\fi
	\expandafter\ifx\csname url\endcsname\relax
	\def\url#1{\texttt{#1}}\fi
	\expandafter\ifx\csname urlprefix\endcsname\relax\def\urlprefix{URL }\fi
	
	\bibitem[{Antonovics(1976)}]{Antonovics1976}
	\ifbool{MyRefNumbers}{\stepcounter{MyBibCount}\theMyBibCount.\\}{}Antonovics,
	J. (1976).
	\newblock The input from population genetics: ``the new ecological genetics''.
	\newblock \emph{Syst. Bot.}, 1, 233--245.
	
	\bibitem[{Aulsebrook \emph{et~al.}(2020)Aulsebrook, Bertram, Martin,
		Aulsebrook, Brodin, Evans, Hall, O'Bryan, Pask, Tyler \&
		Wong}]{Aulsebrook2020}
	\ifbool{MyRefNumbers}{\stepcounter{MyBibCount}\theMyBibCount.\\}{}Aulsebrook,
	L.~C., Bertram, M.~G., Martin, J.~M., Aulsebrook, A.~E., Brodin, T., Evans,
	J.~P., Hall, M.~D., O'Bryan, M.~K., Pask, A.~J., Tyler, C.~R. \& Wong, B.
	B.~M. (2020).
	\newblock Reproduction in a polluted world: implications for wildlife.
	\newblock \emph{Reproduction}, 160, R13--R23.
	
	\bibitem[{Bashforth \emph{et~al.}(2011)Bashforth, Dr{\'{a}}bkov{\'{a}},
		Oplu{\v{s}}til, Gibling \& Falcon-Lang}]{Bashforth2011}
	\ifbool{MyRefNumbers}{\stepcounter{MyBibCount}\theMyBibCount.\\}{}Bashforth,
	A.~R., Dr{\'{a}}bkov{\'{a}}, J., Oplu{\v{s}}til, S., Gibling, M.~R. \&
	Falcon-Lang, H.~J. (2011).
	\newblock Landscape gradients and patchiness in riparian vegetation on a middle
	pennsylvanian braided-river plain prone to flood disturbance
	({N{\'{y}}{\v{r}}any Member, Central and Western Bohemian Basin, Czech
		Republic}).
	\newblock \emph{Rev. Palaeobot. Palyno.}, 163, 153--189.
	
	\bibitem[{Bassar \emph{et~al.}(2021)Bassar, Coulson, Travis \&
		Reznick}]{Bassar2021}
	\ifbool{MyRefNumbers}{\stepcounter{MyBibCount}\theMyBibCount.\\}{}Bassar,
	R.~D., Coulson, T., Travis, J. \& Reznick, D.~N. (2021).
	\newblock Towards a more precise -- and accurate -- view of eco-evolution.
	\newblock \emph{Ecol. Lett.}, 24, 623--625.
	
	\bibitem[{Bassar \emph{et~al.}(2012)Bassar, Ferriere, L{\'{o}}pez-Sepulcre,
		Marshall, Travis, Pringle \& Reznick}]{Bassar2012}
	\ifbool{MyRefNumbers}{\stepcounter{MyBibCount}\theMyBibCount.\\}{}Bassar,
	R.~D., Ferriere, R., L{\'{o}}pez-Sepulcre, A., Marshall, M.~C., Travis, J.,
	Pringle, C.~M. \& Reznick, D.~N. (2012).
	\newblock Direct and indirect ecosystem effects of evolutionary adaptation in
	the {T}rinidadian {G}uppy (\textit{Poecilia reticulata}).
	\newblock \emph{Am. Nat.}, 180, 167--185.
	
	\bibitem[{Bassar \emph{et~al.}(2010)Bassar, Marshall, L{\'{o}}pez-Sepulcre,
		Zandon{\`{a}}, Auer, Travis, Pringle, Flecker, Thomas, Fraser \&
		Reznick}]{Bassar2010}
	\ifbool{MyRefNumbers}{\stepcounter{MyBibCount}\theMyBibCount.\\}{}Bassar,
	R.~D., Marshall, M.~C., L{\'{o}}pez-Sepulcre, A., Zandon{\`{a}}, E., Auer,
	S.~K., Travis, J., Pringle, C.~M., Flecker, A.~S., Thomas, S.~A., Fraser,
	D.~F. \& Reznick, D.~N. (2010).
	\newblock Local adaptation in {T}rinidadian guppies alters ecosystem processes.
	\newblock \emph{Proc. Natl. Acad. Sci. U. S. A.}, 107, 3616--3621.
	
	\bibitem[{Becks \emph{et~al.}(2005)Becks, Hilker, Malchow, Jürgens \&
		Arndt}]{Becks2005}
	\ifbool{MyRefNumbers}{\stepcounter{MyBibCount}\theMyBibCount.\\}{}Becks, L.,
	Hilker, F.~M., Malchow, H., Jürgens, K. \& Arndt, H. (2005).
	\newblock Experimental demonstration of chaos in a microbial food web.
	\newblock \emph{Nature}, 435, 1226--1229.
	
	\bibitem[{Bedau(2008)}]{Bedau2008}
	\ifbool{MyRefNumbers}{\stepcounter{MyBibCount}\theMyBibCount.\\}{}Bedau, M.~A.
	(2008).
	\newblock Is weak emergence just in the mind?
	\newblock \emph{Mind Mach}, 18, 443--459.
	
	\bibitem[{Bernhardt \emph{et~al.}(2018)Bernhardt, Sunday \&
		O'Connor}]{Bernhardt2018}
	\ifbool{MyRefNumbers}{\stepcounter{MyBibCount}\theMyBibCount.\\}{}Bernhardt,
	J.~R., Sunday, J.~M. \& O'Connor, M.~I. (2018).
	\newblock Metabolic theory and the temperature-size rule explain the
	temperature dependence of population carrying capacity.
	\newblock \emph{Am. Nat.}, 192, 687--697.
	
	\bibitem[{Bertness \& Callaway(1994)}]{Bertness1994}
	\ifbool{MyRefNumbers}{\stepcounter{MyBibCount}\theMyBibCount.\\}{}Bertness,
	M.~D. \& Callaway, R. (1994).
	\newblock Positive interactions in communities.
	\newblock \emph{Trends Ecol. Evol.}, 9, 191--193.
	
	\bibitem[{Bickham(2011)}]{Bickham2011}
	\ifbool{MyRefNumbers}{\stepcounter{MyBibCount}\theMyBibCount.\\}{}Bickham,
	J.~W. (2011).
	\newblock The four cornerstones of evolutionary toxicology.
	\newblock \emph{Ecotoxicology}, 20, 497--502.
	
	\bibitem[{Blount \emph{et~al.}(2008)Blount, Borland \& Lenski}]{Blount2008}
	\ifbool{MyRefNumbers}{\stepcounter{MyBibCount}\theMyBibCount.\\}{}Blount,
	Z.~D., Borland, C.~Z. \& Lenski, R.~E. (2008).
	\newblock Historical contingency and the evolution of a key innovation in an
	experimental population of \textit{Escherichia coli}.
	\newblock \emph{Proc. Natl. Acad. Sci. U. S. A.}, 105, 7899--7906.
	
	\bibitem[{Bowler \& Benton(2005)}]{Bowler2005}
	\ifbool{MyRefNumbers}{\stepcounter{MyBibCount}\theMyBibCount.\\}{}Bowler, D.~E.
	\& Benton, T.~G. (2005).
	\newblock Causes and consequences of animal dispersal strategies: relating
	individual behaviour to spatial dynamics.
	\newblock \emph{Biol. Rev.}, 80, 205--225.
	
	\bibitem[{Boyce \emph{et~al.}(2006)Boyce, Haridas, Lee \& {the NCEAS Stochastic
			Demography Working Group}}]{Boyce2006}
	\ifbool{MyRefNumbers}{\stepcounter{MyBibCount}\theMyBibCount.\\}{}Boyce, M.,
	Haridas, C., Lee, C. \& {the NCEAS Stochastic Demography Working Group}
	(2006).
	\newblock Demography in an increasingly variable world.
	\newblock \emph{Trends Ecol. Evol.}, 21, 141--148.
	
	\bibitem[{Brown \emph{et~al.}(2004)Brown, Gillooly, Allen, Savage \&
		West}]{Brown2004}
	\ifbool{MyRefNumbers}{\stepcounter{MyBibCount}\theMyBibCount.\\}{}Brown, J.~H.,
	Gillooly, J.~F., Allen, A.~P., Savage, V.~M. \& West, G.~B. (2004).
	\newblock Toward a metabolic theory of ecology.
	\newblock \emph{Ecology}, 85, 1771--1789.
	
	\bibitem[{Bullock \emph{et~al.}(2018)Bullock, Bonte, Pufal, da~Silva~Carvalho,
		Chapman, Garc{\'{\i}}a, Garc{\'{\i}}a, Matthysen \& Delgado}]{Bullock2018}
	\ifbool{MyRefNumbers}{\stepcounter{MyBibCount}\theMyBibCount.\\}{}Bullock,
	J.~M., Bonte, D., Pufal, G., da~Silva~Carvalho, C., Chapman, D.~S.,
	Garc{\'{\i}}a, C., Garc{\'{\i}}a, D., Matthysen, E. \& Delgado, M.~M. (2018).
	\newblock Human-mediated dispersal and the rewiring of spatial networks.
	\newblock \emph{Trends Ecol. Evol.}, 33, 958--970.
	
	\bibitem[{Burke \& Bonduriansky(2017)}]{Burke2017}
	\ifbool{MyRefNumbers}{\stepcounter{MyBibCount}\theMyBibCount.\\}{}Burke, N.~W.
	\& Bonduriansky, R. (2017).
	\newblock Sexual conflict, facultative asexuality, and the true paradox of sex.
	\newblock \emph{Trends Ecol. Evol.}, 32, 646--652.
	
	\bibitem[{Butterfield(2007)}]{Butterfield2007}
	\ifbool{MyRefNumbers}{\stepcounter{MyBibCount}\theMyBibCount.\\}{}Butterfield,
	N.~J. (2007).
	\newblock Macroevolution and macroecology through deep time.
	\newblock \emph{Palaeontology}, 50, 41--55.
	
	\bibitem[{Butterfield(2011)}]{Butterfield2011}
	\ifbool{MyRefNumbers}{\stepcounter{MyBibCount}\theMyBibCount.\\}{}Butterfield,
	N.~J. (2011).
	\newblock Animals and the invention of the phanerozoic earth system.
	\newblock \emph{Trends Ecol. Evol.}, 26, 81--87.
	
	\bibitem[{Butterfield(2017)}]{Butterfield2017}
	\ifbool{MyRefNumbers}{\stepcounter{MyBibCount}\theMyBibCount.\\}{}Butterfield,
	N.~J. (2017).
	\newblock Oxygen, animals and aquatic bioturbation: An updated account.
	\newblock \emph{Geobiology}, 16, 3--16.
	
	\bibitem[{Cadotte(2007)}]{Cadotte2007}
	\ifbool{MyRefNumbers}{\stepcounter{MyBibCount}\theMyBibCount.\\}{}Cadotte,
	M.~W. (2007).
	\newblock Competition-colonization trade-offs {and} disturbance effects at
	multiple {sc}ales.
	\newblock \emph{Ecology}, 88, 823--829.
	
	\bibitem[{Cadotte \emph{et~al.}(2006)Cadotte, Mai, Jantz, Collins, Keele \&
		Drake}]{Cadotte2006}
	\ifbool{MyRefNumbers}{\stepcounter{MyBibCount}\theMyBibCount.\\}{}Cadotte,
	M.~W., Mai, D.~V., Jantz, S., Collins, M.~D., Keele, M. \& Drake, J.~A.
	(2006).
	\newblock On testing the competition-colonization trade-off in a multispecies
	assemblage.
	\newblock \emph{Am. Nat.}, 168, 704--709.
	
	\bibitem[{Chakravarti \& van Oppen(2018)}]{Chakravarti2018}
	\ifbool{MyRefNumbers}{\stepcounter{MyBibCount}\theMyBibCount.\\}{}Chakravarti,
	L.~J. \& van Oppen, M. J.~H. (2018).
	\newblock Experimental evolution in coral photosymbionts as a tool to increase
	thermal tolerance.
	\newblock \emph{Front. Mar. Sci.}, 5, 227.
	
	\bibitem[{Chitty(1967)}]{Chitty1967}
	\ifbool{MyRefNumbers}{\stepcounter{MyBibCount}\theMyBibCount.\\}{}Chitty, D.
	(1967).
	\newblock The natural selection of self-regulatory behaviour in animal
	populations.
	\newblock \emph{Proc. Ecol. Soc. Aust.}, 2, 51--78.
	
	\bibitem[{Ciliberti \emph{et~al.}(2007)Ciliberti, Martin \&
		Wagner}]{Ciliberti2007}
	\ifbool{MyRefNumbers}{\stepcounter{MyBibCount}\theMyBibCount.\\}{}Ciliberti,
	O., Martin, C. \& Wagner, A. (2007).
	\newblock Innovation and robustness in complex regulatory gene networks.
	\newblock \emph{Proc. Natl. Acad. Sci. U.S.A.}, 104, 13591--13596.
	
	\bibitem[{Collins \& Gardner(2009)}]{Collins2009}
	\ifbool{MyRefNumbers}{\stepcounter{MyBibCount}\theMyBibCount.\\}{}Collins, S.
	\& Gardner, A. (2009).
	\newblock Integrating physiological, ecological and evolutionary change: a
	price equation approach.
	\newblock \emph{Ecol. Lett.}, 12, 744--757.
	
	\bibitem[{Colton \emph{et~al.}(2022)Colton, McManus, Schindler, Mumby, Palumbi,
		Webster, Essington, Fox, Forrest, Schill, Pollock, DeFilippo, Tekwa,
		Walsworth \& Pinsky}]{Colton2022}
	\ifbool{MyRefNumbers}{\stepcounter{MyBibCount}\theMyBibCount.\\}{}Colton,
	M.~A., McManus, L.~C., Schindler, D.~E., Mumby, P.~J., Palumbi, S.~R.,
	Webster, M.~M., Essington, T.~E., Fox, H.~E., Forrest, D.~L., Schill, S.~R.,
	Pollock, F.~J., DeFilippo, L.~B., Tekwa, E.~W., Walsworth, T.~E. \& Pinsky,
	M.~L. (2022).
	\newblock Coral conservation in a warming world must harness evolutionary
	adaptation.
	\newblock \emph{Nat. Ecol. Evol.}, 6, 1405--1407.
	
	\bibitem[{Corenblit \emph{et~al.}(2011)Corenblit, Baas, Bornette, Darrozes,
		Delmotte, Francis, Gurnell, Julien, Naiman \& Steiger}]{Corenblit2011}
	\ifbool{MyRefNumbers}{\stepcounter{MyBibCount}\theMyBibCount.\\}{}Corenblit,
	D., Baas, A.~C., Bornette, G., Darrozes, J., Delmotte, S., Francis, R.~A.,
	Gurnell, A.~M., Julien, F., Naiman, R.~J. \& Steiger, J. (2011).
	\newblock Feedbacks between geomorphology and biota controlling earth surface
	processes and landforms: A review of foundation concepts and current
	understandings.
	\newblock \emph{Earth Sci. Rev.}, 106, 307--331.
	
	\bibitem[{Corenblit \emph{et~al.}(2015)Corenblit, Davies, Steiger, Gibling \&
		Bornette}]{Corenblit2015}
	\ifbool{MyRefNumbers}{\stepcounter{MyBibCount}\theMyBibCount.\\}{}Corenblit,
	D., Davies, N.~S., Steiger, J., Gibling, M.~R. \& Bornette, G. (2015).
	\newblock Considering river structure and stability in the light of evolution:
	feedbacks between riparian vegetation and hydrogeomorphology.
	\newblock \emph{Earth Surf Proc Land}, 40, 189--207.
	
	\bibitem[{Corenblit \emph{et~al.}(2009)Corenblit, Steiger, Gurnell, Tabacchi \&
		Roques}]{Corenblit2009}
	\ifbool{MyRefNumbers}{\stepcounter{MyBibCount}\theMyBibCount.\\}{}Corenblit,
	D., Steiger, J., Gurnell, A.~M., Tabacchi, E. \& Roques, L. (2009).
	\newblock Control of sediment dynamics by vegetation as a key function driving
	biogeomorphic succession within fluvial corridors.
	\newblock \emph{Earth Surf Proc Land}, 34, 1790--1810.
	
	\bibitem[{Cote \emph{et~al.}(accepted)Cote, Dahirel, Schtickzelle, Altermatt,
		Ansart, Blanchet, Chaine, {De Laender}, {De Raedt}, Haegeman, Jacob, Kaltz,
		Laurent, Little, Madec, Manzi, Masier, Pellerin, Pennekamp, Therry, Vong,
		Winandy, Bonte, Fronhofer \& Legrand}]{Coteaccepted}
	\ifbool{MyRefNumbers}{\stepcounter{MyBibCount}\theMyBibCount.\\}{}Cote, J.,
	Dahirel, M., Schtickzelle, N., Altermatt, F., Ansart, A., Blanchet, S.,
	Chaine, A., {De Laender}, F., {De Raedt}, J., Haegeman, B., Jacob, S., Kaltz,
	O., Laurent, E., Little, C.~J., Madec, L., Manzi, F., Masier, S., Pellerin,
	F., Pennekamp, F., Therry, L., Vong, A., Winandy, L., Bonte, D., Fronhofer,
	E.~A. \& Legrand, D. (accepted).
	\newblock Dispersal syndromes in challenging environments: a cross-species
	experiment.
	\newblock \emph{Ecol. Lett.}
	
	\bibitem[{Covich(2010)}]{Covich2010}
	\ifbool{MyRefNumbers}{\stepcounter{MyBibCount}\theMyBibCount.\\}{}Covich, A.~P.
	(2010).
	\newblock Winning the biodiversity arms race among freshwater gastropods:
	competition and coexistence through shell variability and predator avoidance.
	\newblock \emph{Hydrobiologia}, 653, 191--215.
	
	\bibitem[{Davies \emph{et~al.}(2021)Davies, Berry, Marshall, Wellman \&
		Lindemann}]{Davies2021}
	\ifbool{MyRefNumbers}{\stepcounter{MyBibCount}\theMyBibCount.\\}{}Davies,
	N.~S., Berry, C.~M., Marshall, J.~E., Wellman, C.~H. \& Lindemann, F.-J.
	(2021).
	\newblock The {D}evonian landscape factory: plant--sediment interactions in the
	{O}ld {R}ed {S}andstone of {S}valbard and the rise of vegetation as a
	biogeomorphic agent.
	\newblock \emph{J. Geol. Soc. London}, 178, jgs2020--225.
	
	\bibitem[{Davies \& Gibling(2011)}]{Davies2011}
	\ifbool{MyRefNumbers}{\stepcounter{MyBibCount}\theMyBibCount.\\}{}Davies, N.~S.
	\& Gibling, M.~R. (2011).
	\newblock Evolution of fixed-channel alluvial plains in response to
	carboniferous vegetation.
	\newblock \emph{Nat Geosci}, 4, 629--633.
	
	\bibitem[{Davies \& Gibling(2013)}]{Davies2013}
	\ifbool{MyRefNumbers}{\stepcounter{MyBibCount}\theMyBibCount.\\}{}Davies, N.~S.
	\& Gibling, M.~R. (2013).
	\newblock The sedimentary record of carboniferous rivers: Continuing influence
	of land plant evolution on alluvial processes and palaeozoic ecosystems.
	\newblock \emph{Earth Sci. Rev.}, 120, 40--79.
	
	\bibitem[{Dawkins(1982)}]{Dawkins1982}
	\ifbool{MyRefNumbers}{\stepcounter{MyBibCount}\theMyBibCount.\\}{}Dawkins, R.
	(1982).
	\newblock \emph{The Extended Phenotype}.
	\newblock Oxford University Press.
	
	\bibitem[{Dawkins(2004)}]{Dawkins2004}
	\ifbool{MyRefNumbers}{\stepcounter{MyBibCount}\theMyBibCount.\\}{}Dawkins, R.
	(2004).
	\newblock Extended phenotype --- but not too extended. a reply to laland,
	turner and jablonka.
	\newblock \emph{Biol. Philos.}, 19, 377--396.
	
	\bibitem[{Dell \emph{et~al.}(2014)Dell, Pawar \& Savage}]{Dell2014a}
	\ifbool{MyRefNumbers}{\stepcounter{MyBibCount}\theMyBibCount.\\}{}Dell, A.~I.,
	Pawar, S. \& Savage, V.~M. (2014).
	\newblock Temperature dependence of trophic interactions are driven by
	asymmetry of species responses and foraging strategy.
	\newblock \emph{J. Anim. Ecol.}, 83, 70--84.
	
	\bibitem[{DeLong \emph{et~al.}(2016)DeLong, Forbes, Galic, Gibert, Laport,
		Phillips \& Vavra}]{DeLong2016}
	\ifbool{MyRefNumbers}{\stepcounter{MyBibCount}\theMyBibCount.\\}{}DeLong,
	J.~P., Forbes, V.~E., Galic, N., Gibert, J.~P., Laport, R.~G., Phillips,
	J.~S. \& Vavra, J.~M. (2016).
	\newblock How fast is fast? {Eco-evolutionary} dynamics and rates of change in
	populations and phenotypes.
	\newblock \emph{Ecol. Evol.}, 6, 573--581.
	
	\bibitem[{Deshpande \& Fronhofer(2022)}]{Deshpande2022}
	\ifbool{MyRefNumbers}{\stepcounter{MyBibCount}\theMyBibCount.\\}{}Deshpande,
	J.~N. \& Fronhofer, E.~A. (2022).
	\newblock Genetic architecture of dispersal and local adaptation drives
	accelerating range expansions.
	\newblock \emph{Proc. Natl. Acad. Sci. U. S. A.}, 119, e2121858119.
	
	\bibitem[{DiMichele \emph{et~al.}(2005)DiMichele, Gastaldo \&
		Pfefferkorn}]{DiMichele2005}
	\ifbool{MyRefNumbers}{\stepcounter{MyBibCount}\theMyBibCount.\\}{}DiMichele,
	W.~A., Gastaldo, R.~A. \& Pfefferkorn, H.~W. (2005).
	\newblock Plant biodiversity partitioning in the late carboniferous and early
	permian and its implications for ecosystem assembly.
	\newblock \emph{Proc. Calif. Acad. Sci.}, 56, 32--49.
	
	\bibitem[{Drake \emph{et~al.}(2020)Drake, O’Regan, Dakos, Kéfi \&
		Rohani.}]{Drake2020}
	\ifbool{MyRefNumbers}{\stepcounter{MyBibCount}\theMyBibCount.\\}{}Drake, J.~M.,
	O’Regan, S.~M., Dakos, V., Kéfi, S. \& Rohani., P. (2020).
	\newblock Alternative stable states, tipping points, and early warning signals
	of ecological transitions.
	\newblock In: \emph{Theoretical Ecology: concepts and applications} (eds.
	McCann, K.~S. \& Gellner, G.). Oxford University Press.
	
	\bibitem[{El-Sabaawi \emph{et~al.}(2014)El-Sabaawi, Marshall, Bassar,
		L{\'{o}}pez-Sepulcre, Palkovacs \& Dalton}]{ElSabaawi2014}
	\ifbool{MyRefNumbers}{\stepcounter{MyBibCount}\theMyBibCount.\\}{}El-Sabaawi,
	R.~W., Marshall, M.~C., Bassar, R.~D., L{\'{o}}pez-Sepulcre, A., Palkovacs,
	E.~P. \& Dalton, C. (2014).
	\newblock Assessing the effects of guppy life history evolution on nutrient
	recycling: from experiments to the field.
	\newblock \emph{Freshwater Biol.}, 60, 590--601.
	
	\bibitem[{Faillace \emph{et~al.}(2021)Faillace, Sentis \&
		Montoya}]{Faillace2021}
	\ifbool{MyRefNumbers}{\stepcounter{MyBibCount}\theMyBibCount.\\}{}Faillace,
	C.~A., Sentis, A. \& Montoya, J.~M. (2021).
	\newblock Eco-evolutionary consequences of habitat warming and fragmentation in
	communities.
	\newblock \emph{Biol. Rev.}, 96, 1933--1950.
	
	\bibitem[{Falcon-Lang \emph{et~al.}(2011)Falcon-Lang, Jud, Nelson, DiMichele,
		Chaney \& Lucas}]{FalconLang2011}
	\ifbool{MyRefNumbers}{\stepcounter{MyBibCount}\theMyBibCount.\\}{}Falcon-Lang,
	H.~J., Jud, N.~A., Nelson, W.~J., DiMichele, W.~A., Chaney, D.~S. \& Lucas,
	S.~G. (2011).
	\newblock Pennsylvanian coniferopsid forests in sabkha facies reveal the nature
	of seasonal tropical biome.
	\newblock \emph{Geology}, 39, 371--374.
	
	\bibitem[{Fronhofer \emph{et~al.}(2020)Fronhofer, Govaert, O'Connor, Schreiber
		\& Altermatt}]{Fronhofersubmitted}
	\ifbool{MyRefNumbers}{\stepcounter{MyBibCount}\theMyBibCount.\\}{}Fronhofer,
	E.~A., Govaert, L., O'Connor, M.~I., Schreiber, S.~J. \& Altermatt, F.
	(2020).
	\newblock The shape of density dependence and the relationship between
	population growth, intraspecific competition and equilibrium population
	density.
	\newblock \emph{bioRxiv --- DOI: 10.1101/485946}.
	
	\bibitem[{Fronhofer \emph{et~al.}(2018)Fronhofer, Legrand, Altermatt, Ansart,
		Blanchet, Bonte, Chaine, Dahirel, {De Laender}, {De Raedt}, di~Gesu, Jacob,
		Kaltz, Laurent, Little, Madec, Manzi, Masier, Pellerin, Pennekamp,
		Schtickzelle, Therry, Vong, Winandy \& Cote}]{Fronhofer2018}
	\ifbool{MyRefNumbers}{\stepcounter{MyBibCount}\theMyBibCount.\\}{}Fronhofer,
	E.~A., Legrand, D., Altermatt, F., Ansart, A., Blanchet, S., Bonte, D.,
	Chaine, A., Dahirel, M., {De Laender}, F., {De Raedt}, J., di~Gesu, L.,
	Jacob, S., Kaltz, O., Laurent, E., Little, C.~J., Madec, L., Manzi, F.,
	Masier, S., Pellerin, F., Pennekamp, F., Schtickzelle, N., Therry, L., Vong,
	A., Winandy, L. \& Cote, J. (2018).
	\newblock Bottom-up and top-down control of dispersal across major organismal
	groups.
	\newblock \emph{Nat. Ecol. Evol.}, 2, 1859--1863.
	
	\bibitem[{Fussmann \emph{et~al.}(2007)Fussmann, Loreau \&
		Abrams}]{Fussmann2007}
	\ifbool{MyRefNumbers}{\stepcounter{MyBibCount}\theMyBibCount.\\}{}Fussmann,
	G.~F., Loreau, M. \& Abrams, P.~A. (2007).
	\newblock Eco-evolutionary dynamics of communities and ecosystems.
	\newblock \emph{Funct. Ecol.}, 21, 465--477.
	
	\bibitem[{Futuyma(2010)}]{Futuyma2010}
	\ifbool{MyRefNumbers}{\stepcounter{MyBibCount}\theMyBibCount.\\}{}Futuyma,
	D.~J. (2010).
	\newblock Evolutionary constraint and ecological consequences.
	\newblock \emph{Evolution}, 64, 1865--1884.
	
	\bibitem[{Gerber \emph{et~al.}(2018)Gerber, Kokko, Ebert \&
		Booksmythe}]{Gerber2018}
	\ifbool{MyRefNumbers}{\stepcounter{MyBibCount}\theMyBibCount.\\}{}Gerber, N.,
	Kokko, H., Ebert, D. \& Booksmythe, I. (2018).
	\newblock \textit{Daphnia} invest in sexual reproduction when its relative
	costs are reduced.
	\newblock \emph{Proc. R. Soc. B-Biol. Sci.}, 285, 20172176.
	
	\bibitem[{Gibling \emph{et~al.}(2014)Gibling, Davies, Falcon-Lang, Bashforth,
		DiMichele, Rygel \& Ielpi}]{Gibling2014}
	\ifbool{MyRefNumbers}{\stepcounter{MyBibCount}\theMyBibCount.\\}{}Gibling, M.,
	Davies, N., Falcon-Lang, H., Bashforth, A., DiMichele, W., Rygel, M. \&
	Ielpi, A. (2014).
	\newblock Palaeozoic co-evolution of rivers and vegetation: a synthesis of
	current knowledge.
	\newblock \emph{Proceedings of the Geologists Association}, 125, 524--533.
	
	\bibitem[{Gibling \& Davies(2012)}]{Gibling2012}
	\ifbool{MyRefNumbers}{\stepcounter{MyBibCount}\theMyBibCount.\\}{}Gibling,
	M.~R. \& Davies, N.~S. (2012).
	\newblock Palaeozoic landscapes shaped by plant evolution.
	\newblock \emph{Nat Geosci}, 5, 99--105.
	
	\bibitem[{Gilbert \emph{et~al.}(2014)Gilbert, Tunney, McCann, DeLong, Vasseur,
		Savage, Shurin, Dell, Barton, Harley, Kharouba, Kratina, Blanchard, Clements,
		Winder, Greig \& O'Connor}]{Gilbert2014}
	\ifbool{MyRefNumbers}{\stepcounter{MyBibCount}\theMyBibCount.\\}{}Gilbert, B.,
	Tunney, T., McCann, K., DeLong, J., Vasseur, D., Savage, V., Shurin, J.~B.,
	Dell, A.~I., Barton, B.~T., Harley, C. D.~G., Kharouba, H.~M., Kratina, P.,
	Blanchard, J.~L., Clements, C., Winder, M., Greig, H.~S. \& O'Connor, M.~I.
	(2014).
	\newblock A bioenergetic framework for the temperature dependence of trophic
	interactions.
	\newblock \emph{Ecol. Lett.}, 17, 902--914.
	
	\bibitem[{Gillespie \emph{et~al.}(2012)Gillespie, Baldwin, Waters, Fraser,
		Nikula \& Roderick}]{Gillespie2012}
	\ifbool{MyRefNumbers}{\stepcounter{MyBibCount}\theMyBibCount.\\}{}Gillespie,
	R.~G., Baldwin, B.~G., Waters, J.~M., Fraser, C.~I., Nikula, R. \& Roderick,
	G.~K. (2012).
	\newblock Long-distance dispersal: a framework for hypothesis testing.
	\newblock \emph{Trends Ecol. Evol.}, 27, 47--56.
	
	\bibitem[{Gillooly \emph{et~al.}(2001)Gillooly, Brown, West, Savage \&
		Charnov}]{Gillooly2001}
	\ifbool{MyRefNumbers}{\stepcounter{MyBibCount}\theMyBibCount.\\}{}Gillooly,
	J.~F., Brown, J.~H., West, G.~B., Savage, V.~M. \& Charnov, E.~L. (2001).
	\newblock Effects of size and temperature on metabolic rate.
	\newblock \emph{Science}, 293, 2248--2251.
	
	\bibitem[{Gingerich(2001)}]{Gingerich2001}
	\ifbool{MyRefNumbers}{\stepcounter{MyBibCount}\theMyBibCount.\\}{}Gingerich,
	P.~D. (2001).
	\newblock Rates of evolution on the time scale of the evolutionary process.
	\newblock \emph{Genetica}, 112--113, 127--144.
	
	\bibitem[{Gingerich(2009)}]{Gingerich2009}
	\ifbool{MyRefNumbers}{\stepcounter{MyBibCount}\theMyBibCount.\\}{}Gingerich,
	P.~D. (2009).
	\newblock Rates of evolution.
	\newblock \emph{Annu. Rev. Ecol. Evol. Syst.}, 40, 657--675.
	
	\bibitem[{Giraud(2001)}]{Giraud2001}
	\ifbool{MyRefNumbers}{\stepcounter{MyBibCount}\theMyBibCount.\\}{}Giraud, A.
	(2001).
	\newblock The rise and fall of mutator bacteria.
	\newblock \emph{Curr Opin Microbiol}, 4, 582--585.
	
	\bibitem[{Gl{\'{e}}min(2003)}]{Glemin2003}
	\ifbool{MyRefNumbers}{\stepcounter{MyBibCount}\theMyBibCount.\\}{}Gl{\'{e}}min,
	S. (2003).
	\newblock How are deleterious mutations purged? {D}rift versus nonrandom
	mating.
	\newblock \emph{Evolution}, 57, 2678--2687.
	
	\bibitem[{Gould(1989)}]{Gould1989}
	\ifbool{MyRefNumbers}{\stepcounter{MyBibCount}\theMyBibCount.\\}{}Gould, S.~J.
	(1989).
	\newblock \emph{Wonderful Life: The Burgess Shale and the Nature of History}.
	\newblock W. W. Norton \& Co.
	
	\bibitem[{Gounand \emph{et~al.}(2018{\natexlab{a}})Gounand, Harvey, Little \&
		Altermatt}]{Gounand2018}
	\ifbool{MyRefNumbers}{\stepcounter{MyBibCount}\theMyBibCount.\\}{}Gounand, I.,
	Harvey, E., Little, C.~J. \& Altermatt, F. (2018{\natexlab{a}}).
	\newblock Meta-ecosystems 2.0: Rooting the theory into the field.
	\newblock \emph{Trends. Ecol. Evol.}, 33, 36--46.
	
	\bibitem[{Gounand \emph{et~al.}(2018{\natexlab{b}})Gounand, Little, Harvey \&
		Altermatt}]{Gounand2018a}
	\ifbool{MyRefNumbers}{\stepcounter{MyBibCount}\theMyBibCount.\\}{}Gounand, I.,
	Little, C.~J., Harvey, E. \& Altermatt, F. (2018{\natexlab{b}}).
	\newblock Cross-ecosystem carbon flows connecting ecosystems worldwide.
	\newblock \emph{Nat. Commun.}, 9, 4825.
	
	\bibitem[{Govaert \emph{et~al.}(2019)Govaert, Fronhofer, Lion, Eizaguirre,
		Bonte, Egas, Hendry, Martins, Meli\'an, Raeymaekers, Ratikainen, Saether,
		Schweitzer \& Matthews}]{Govaert2019}
	\ifbool{MyRefNumbers}{\stepcounter{MyBibCount}\theMyBibCount.\\}{}Govaert, L.,
	Fronhofer, E.~A., Lion, S., Eizaguirre, C., Bonte, D., Egas, M., Hendry,
	A.~P., Martins, A. D.~B., Meli\'an, C.~J., Raeymaekers, J., Ratikainen,
	I.~I., Saether, B.-E., Schweitzer, J.~A. \& Matthews, B. (2019).
	\newblock Eco-evolutionary feedbacks --- theoretical models and perspectives.
	\newblock \emph{Funct. Ecol.}, 33, 13--30.
	
	\bibitem[{Govaert \emph{et~al.}(2016)Govaert, Pantel \& Meester}]{Govaert2016}
	\ifbool{MyRefNumbers}{\stepcounter{MyBibCount}\theMyBibCount.\\}{}Govaert, L.,
	Pantel, J.~H. \& Meester, L.~D. (2016).
	\newblock Eco-evolutionary partitioning metrics: assessing the importance of
	ecological and evolutionary contributions to population and community change.
	\newblock \emph{Ecol. Lett.}, 19, 839--853.
	
	\bibitem[{Grantham(2007)}]{Grantham2007}
	\ifbool{MyRefNumbers}{\stepcounter{MyBibCount}\theMyBibCount.\\}{}Grantham, T.
	(2007).
	\newblock Is macroevolution more than successive rounds of microevolution?
	\newblock \emph{Palaeontology}, 50, 75--85.
	
	\bibitem[{Greb \emph{et~al.}(2006)Greb, DiMichele \& Gastaldo}]{Greb2006}
	\ifbool{MyRefNumbers}{\stepcounter{MyBibCount}\theMyBibCount.\\}{}Greb, S.~F.,
	DiMichele, W.~A. \& Gastaldo, R.~A. (2006).
	\newblock Evolution and importance of wetlands in earth history.
	\newblock In: \emph{Wetlands through Time} (eds. Greb, S.~F. \& DiMichele,
	W.~A.), vol. 399.
	
	\bibitem[{Gurnell(2014)}]{Gurnell2014}
	\ifbool{MyRefNumbers}{\stepcounter{MyBibCount}\theMyBibCount.\\}{}Gurnell, A.
	(2014).
	\newblock Plants as river system engineers.
	\newblock \emph{Earth Surf Proc Land}, 39, 4--25.
	
	\bibitem[{Hairston \emph{et~al.}(2005)Hairston, Ellner, Geber, Yoshida \&
		Fox}]{Hairston2005}
	\ifbool{MyRefNumbers}{\stepcounter{MyBibCount}\theMyBibCount.\\}{}Hairston,
	N.~G., Ellner, S.~P., Geber, M.~A., Yoshida, T. \& Fox, J.~A. (2005).
	\newblock Rapid evolution and the convergence of ecological and evolutionary
	time.
	\newblock \emph{Ecol. Lett.}, 8, 1114--1127.
	
	\bibitem[{Hamann \emph{et~al.}(2020)Hamann, Blevins, Franks, Jameel \&
		Anderson}]{Hamann2020}
	\ifbool{MyRefNumbers}{\stepcounter{MyBibCount}\theMyBibCount.\\}{}Hamann, E.,
	Blevins, C., Franks, S.~J., Jameel, M.~I. \& Anderson, J.~T. (2020).
	\newblock Climate change alters plant-herbivore interactions.
	\newblock \emph{New Phytol.}, 229, 1894--1910.
	
	\bibitem[{Hassell(1975)}]{Hassell1975}
	\ifbool{MyRefNumbers}{\stepcounter{MyBibCount}\theMyBibCount.\\}{}Hassell,
	M.~P. (1975).
	\newblock Density-dependence in single-species populations.
	\newblock \emph{J. Anim. Ecol.}, 44, 283--295.
	
	\bibitem[{Hassell \emph{et~al.}(1976)Hassell, Lawton \& May}]{Hassell1976}
	\ifbool{MyRefNumbers}{\stepcounter{MyBibCount}\theMyBibCount.\\}{}Hassell,
	M.~P., Lawton, J.~H. \& May, R.~M. (1976).
	\newblock Patterns of dynamical behavior in single-species populations.
	\newblock \emph{J. Anim. Ecol.}, 45, 471--486.
	
	\bibitem[{Hendry(2017)}]{Hendry2017}
	\ifbool{MyRefNumbers}{\stepcounter{MyBibCount}\theMyBibCount.\\}{}Hendry, A.~P.
	(2017).
	\newblock \emph{Eco-evolutionary dynamics}.
	\newblock Princeton University Press, Princeton, NJ, USA.
	
	\bibitem[{Hendry \& Kinnison(1999)}]{Hendry1999}
	\ifbool{MyRefNumbers}{\stepcounter{MyBibCount}\theMyBibCount.\\}{}Hendry, A.~P.
	\& Kinnison, M.~T. (1999).
	\newblock The pace of modern life: measuring rates of contemporary
	microevolution.
	\newblock \emph{Evolution}, 53, 1637--1653.
	
	\bibitem[{Hiltunen \emph{et~al.}(2014)Hiltunen, Hairston, Hooker, Jones \&
		Ellner}]{Hiltunen2014}
	\ifbool{MyRefNumbers}{\stepcounter{MyBibCount}\theMyBibCount.\\}{}Hiltunen, T.,
	Hairston, N.~G., Hooker, G., Jones, L.~E. \& Ellner, S.~P. (2014).
	\newblock A newly discovered role of evolution in previously published
	consumer-resource dynamics.
	\newblock \emph{Ecol. Lett.}, 17, 915--923.
	
	\bibitem[{Hoffman \emph{et~al.}(1998)Hoffman, Kaufman, Halverson \&
		Schrag}]{Hoffman1998}
	\ifbool{MyRefNumbers}{\stepcounter{MyBibCount}\theMyBibCount.\\}{}Hoffman,
	P.~F., Kaufman, A.~J., Halverson, G.~P. \& Schrag, D.~P. (1998).
	\newblock A neoproterozoic snowball earth.
	\newblock \emph{Science}, 281, 1342--1346.
	
	\bibitem[{Holland(2006)}]{Holland2006}
	\ifbool{MyRefNumbers}{\stepcounter{MyBibCount}\theMyBibCount.\\}{}Holland,
	H.~D. (2006).
	\newblock The oxygenation of the atmosphere and oceans.
	\newblock \emph{Philos. Trans. R. Soc. B-Biol. Sci.}, 361, 903--915.
	
	\bibitem[{Houlahan \emph{et~al.}(2016)Houlahan, McKinney, Anderson \&
		McGill}]{Houlahan2016}
	\ifbool{MyRefNumbers}{\stepcounter{MyBibCount}\theMyBibCount.\\}{}Houlahan,
	J.~E., McKinney, S.~T., Anderson, T.~M. \& McGill, B.~J. (2016).
	\newblock The priority of prediction in ecological understanding.
	\newblock \emph{Oikos}, 126, 1--7.
	
	\bibitem[{Houle(1992)}]{Houle1992}
	\ifbool{MyRefNumbers}{\stepcounter{MyBibCount}\theMyBibCount.\\}{}Houle, D.
	(1992).
	\newblock Comparing evolvability and variability of quantitative traits.
	\newblock \emph{Genetics}, 130, 195--204.
	
	\bibitem[{Hoverman \emph{et~al.}(2005)Hoverman, Auld \& Relyea}]{Hoverman2005}
	\ifbool{MyRefNumbers}{\stepcounter{MyBibCount}\theMyBibCount.\\}{}Hoverman,
	J.~T., Auld, J.~R. \& Relyea, R.~A. (2005).
	\newblock Putting prey back together again: integrating predator-induced
	behavior, morphology, and life history.
	\newblock \emph{Oecologia}, 144, 481--491.
	
	\bibitem[{Huneman(2008)}]{Huneman2008}
	\ifbool{MyRefNumbers}{\stepcounter{MyBibCount}\theMyBibCount.\\}{}Huneman, P.
	(2008).
	\newblock Emergence made ontological? {C}omputational versus combinatorial
	approaches.
	\newblock \emph{Philos. Sci.}, 75, 595--607.
	
	\bibitem[{Huneman(2019)}]{Huneman2019}
	\ifbool{MyRefNumbers}{\stepcounter{MyBibCount}\theMyBibCount.\\}{}Huneman, P.
	(2019).
	\newblock How the modern synthesis came to ecology.
	\newblock \emph{J. Hist. Biol.}, 52, 635--686.
	
	\bibitem[{Hunter(1998)}]{Hunter1998}
	\ifbool{MyRefNumbers}{\stepcounter{MyBibCount}\theMyBibCount.\\}{}Hunter, J.~P.
	(1998).
	\newblock Key innovations and the ecology of macroevolution.
	\newblock \emph{Trends Ecol. Evol.}, 13, 31--36.
	
	\bibitem[{Hutchinson(1965)}]{Hutchinson1965}
	\ifbool{MyRefNumbers}{\stepcounter{MyBibCount}\theMyBibCount.\\}{}Hutchinson,
	G.~E. (1965).
	\newblock \emph{The Ecological Theater and the Evolutionary Play}.
	\newblock Yale University Press.
	
	\bibitem[{Johnson(2002)}]{Johnson2002a}
	\ifbool{MyRefNumbers}{\stepcounter{MyBibCount}\theMyBibCount.\\}{}Johnson,
	D.~L. (2002).
	\newblock Darwin would be proud: Bioturbation, dynamic denudation, and the
	power of theory in science.
	\newblock \emph{Geoarchaeology}, 17, 7--40.
	
	\bibitem[{Jones \emph{et~al.}(1994)Jones, Lawton \& Shachak}]{Jones1994}
	\ifbool{MyRefNumbers}{\stepcounter{MyBibCount}\theMyBibCount.\\}{}Jones, C.~G.,
	Lawton, J.~H. \& Shachak, M. (1994).
	\newblock Organisms as ecosystem engineers.
	\newblock \emph{Oikos}, 69, 373--386.
	
	\bibitem[{Judson(2017)}]{Judson2017}
	\ifbool{MyRefNumbers}{\stepcounter{MyBibCount}\theMyBibCount.\\}{}Judson, O.~P.
	(2017).
	\newblock The energy expansions of evolution.
	\newblock \emph{Nat. Ecol. Evol.}, 1, 0138.
	
	\bibitem[{Kennedy \emph{et~al.}(2012)Kennedy, Gensel \& Gibling}]{Kennedy2012}
	\ifbool{MyRefNumbers}{\stepcounter{MyBibCount}\theMyBibCount.\\}{}Kennedy,
	K.~L., Gensel, P.~G. \& Gibling, M.~R. (2012).
	\newblock Paleoenvironmental inferences from the classic lower devonian
	plant-bearing locality of the campbellton formation, new brunswick, canada.
	\newblock \emph{Palaios}, 27, 424--438.
	
	\bibitem[{Kiessling \emph{et~al.}(1999)Kiessling, Flügel \&
		Golonka}]{Kiessling1999}
	\ifbool{MyRefNumbers}{\stepcounter{MyBibCount}\theMyBibCount.\\}{}Kiessling,
	W., Flügel, E. \& Golonka, J. (1999).
	\newblock Paleoreef maps: Evaluation of a comprehensive database on phanerozoic
	reefs.
	\newblock \emph{AAPG Bulletin}, 83, 1552--1587.
	
	\bibitem[{Kokko \& López-Sepulcre(2007)}]{Kokko2007}
	\ifbool{MyRefNumbers}{\stepcounter{MyBibCount}\theMyBibCount.\\}{}Kokko, H. \&
	López-Sepulcre, A. (2007).
	\newblock The ecogenetic link between demography and evolution: can we bridge
	the gap between theory and data?
	\newblock \emph{Ecol. Lett.}, 10, 773--782.
	
	\bibitem[{Kra{\v{s}}ovec \emph{et~al.}(2017)Kra{\v{s}}ovec, Richards, Gifford,
		Hatcher, Faulkner, Belavkin, Channon, Aston, McBain \& Knight}]{Krasovec2017}
	\ifbool{MyRefNumbers}{\stepcounter{MyBibCount}\theMyBibCount.\\}{}Kra{\v{s}}ovec,
	R., Richards, H., Gifford, D.~R., Hatcher, C., Faulkner, K.~J., Belavkin,
	R.~V., Channon, A., Aston, E., McBain, A.~J. \& Knight, C.~G. (2017).
	\newblock Spontaneous mutation rate is a plastic trait associated with
	population density across domains of life.
	\newblock \emph{PLoS Biol.}, 15, e2002731.
	
	\bibitem[{Kylafis \& Loreau(2008)}]{Kylafis2008}
	\ifbool{MyRefNumbers}{\stepcounter{MyBibCount}\theMyBibCount.\\}{}Kylafis, G.
	\& Loreau, M. (2008).
	\newblock Ecological and evolutionary consequences of niche construction for
	its agent.
	\newblock \emph{Ecol. Lett.}, 11, 1072--1081.
	
	\bibitem[{Labandeira(1998)}]{Labandeira1998}
	\ifbool{MyRefNumbers}{\stepcounter{MyBibCount}\theMyBibCount.\\}{}Labandeira,
	C.~C. (1998).
	\newblock Early history of arthropod and vascular plant associations.
	\newblock \emph{Annu. Rev. Earth Pl. Sc.}, 26, 329--377.
	
	\bibitem[{Lagerstrom \emph{et~al.}(2022)Lagerstrom, Vance, Cornwell, Ruffley,
		Bellagio, Exposito-Alonso, Palumbi \& Hadly}]{Lagerstrom2022}
	\ifbool{MyRefNumbers}{\stepcounter{MyBibCount}\theMyBibCount.\\}{}Lagerstrom,
	K.~M., Vance, S., Cornwell, B.~H., Ruffley, M., Bellagio, T.,
	Exposito-Alonso, M., Palumbi, S.~R. \& Hadly, E.~A. (2022).
	\newblock From coral reefs to joshua trees: What ecological interactions teach
	us about the adaptive capacity of biodiversity in the anthropocene.
	\newblock \emph{Philos. Trans. R. Soc. B-Biol. Sci.}, 377.
	
	\bibitem[{Laland \emph{et~al.}(2016)Laland, Matthews \& Feldman}]{Laland2016}
	\ifbool{MyRefNumbers}{\stepcounter{MyBibCount}\theMyBibCount.\\}{}Laland, K.,
	Matthews, B. \& Feldman, M.~W. (2016).
	\newblock An introduction to niche construction theory.
	\newblock \emph{Evol. Ecol.}, 30, 191--202.
	
	\bibitem[{Laland(2004)}]{Laland2004}
	\ifbool{MyRefNumbers}{\stepcounter{MyBibCount}\theMyBibCount.\\}{}Laland, K.~L.
	(2004).
	\newblock Extending the extended phenotype.
	\newblock \emph{Biol. Philos.}, 19, 313--325.
	
	\bibitem[{Laland \emph{et~al.}(1999)Laland, Odling-Smee \&
		Feldman}]{Laland1999}
	\ifbool{MyRefNumbers}{\stepcounter{MyBibCount}\theMyBibCount.\\}{}Laland,
	K.~N., Odling-Smee, F.~J. \& Feldman, M.~W. (1999).
	\newblock Evolutionary consequences of niche construction and their
	implications for ecology.
	\newblock \emph{Proc. Natl. Acad. Sci. U. S. A.}, 96, 10242--10247.
	
	\bibitem[{Laroche \emph{et~al.}(2016)Laroche, Jarne, Perrot \&
		Massol}]{Laroche2016}
	\ifbool{MyRefNumbers}{\stepcounter{MyBibCount}\theMyBibCount.\\}{}Laroche, F.,
	Jarne, P., Perrot, T. \& Massol, F. (2016).
	\newblock The evolution of the competition{\textendash}dispersal trade-off
	affects $\alpha$- and $\beta$-diversity in a heterogeneous metacommunity.
	\newblock \emph{Proc. R. Soc. B-Biol. Sci.}, 283, 20160548.
	
	\bibitem[{Lawson \emph{et~al.}(2015)Lawson, Vindenes, Bailey \& van~de
		Pol}]{Lawson2015}
	\ifbool{MyRefNumbers}{\stepcounter{MyBibCount}\theMyBibCount.\\}{}Lawson,
	C.~R., Vindenes, Y., Bailey, L. \& van~de Pol, M. (2015).
	\newblock Environmental variation and population responses to global change.
	\newblock \emph{Ecol. Lett.}, 18, 724--736.
	
	\bibitem[{Lee \& Coop(2017)}]{Lee2017}
	\ifbool{MyRefNumbers}{\stepcounter{MyBibCount}\theMyBibCount.\\}{}Lee, K.~M. \&
	Coop, G. (2017).
	\newblock Distinguishing among modes of convergent adaptation using population
	genomic data.
	\newblock \emph{Genetics}, 207, 1591--1619.
	
	\bibitem[{Legrand \emph{et~al.}(2017)Legrand, Cote, Fronhofer, Holt, Ronce,
		Schtickzelle, Travis \& Clobert}]{Legrand2017}
	\ifbool{MyRefNumbers}{\stepcounter{MyBibCount}\theMyBibCount.\\}{}Legrand, D.,
	Cote, J., Fronhofer, E.~A., Holt, R.~D., Ronce, O., Schtickzelle, N., Travis,
	J. M.~J. \& Clobert, J. (2017).
	\newblock Eco-evolutionary dynamics in fragmented landscapes.
	\newblock \emph{Ecography}, 40, 9--25.
	
	\bibitem[{Lehmann(2007)}]{Lehmann2007}
	\ifbool{MyRefNumbers}{\stepcounter{MyBibCount}\theMyBibCount.\\}{}Lehmann, L.
	(2007).
	\newblock The evolution of trans-generational altruism: kin selection meets
	niche construction.
	\newblock \emph{J. Evol. Biol.}, 20, 181--189.
	
	\bibitem[{Lenormand \emph{et~al.}(2009)Lenormand, Roze \&
		Rousset}]{Lenormand2009}
	\ifbool{MyRefNumbers}{\stepcounter{MyBibCount}\theMyBibCount.\\}{}Lenormand,
	T., Roze, D. \& Rousset, F. (2009).
	\newblock Stochasticity in evolution.
	\newblock \emph{Trends Ecol. Evol.}, 24, 157--165.
	
	\bibitem[{Levin(1998)}]{Levin1998}
	\ifbool{MyRefNumbers}{\stepcounter{MyBibCount}\theMyBibCount.\\}{}Levin, S.~A.
	(1998).
	\newblock Ecosystems and the biosphere as complex adaptive systems.
	\newblock \emph{Ecosystems}, 1, 431--436.
	
	\bibitem[{Lion(2018)}]{Lion2018}
	\ifbool{MyRefNumbers}{\stepcounter{MyBibCount}\theMyBibCount.\\}{}Lion, S.
	(2018).
	\newblock Theoretical approaches in evolutionary ecology: Environmental
	feedback as a unifying perspective.
	\newblock \emph{Am. Nat.}, 191, 21--44.
	
	\bibitem[{Longrich \emph{et~al.}(2012)Longrich, Bhullar \&
		Gauthier}]{Longrich2012}
	\ifbool{MyRefNumbers}{\stepcounter{MyBibCount}\theMyBibCount.\\}{}Longrich,
	N.~R., Bhullar, B.-A.~S. \& Gauthier, J.~A. (2012).
	\newblock Mass extinction of lizards and snakes at the cretaceous–paleogene
	boundary.
	\newblock \emph{Proc. Natl. Acad. Sci. U. S. A.}, 109, 21396--21401.
	
	\bibitem[{L{\'{o}}pez-Barea \& Pueyo(1998)}]{LopezBarea1998}
	\ifbool{MyRefNumbers}{\stepcounter{MyBibCount}\theMyBibCount.\\}{}L{\'{o}}pez-Barea,
	J. \& Pueyo, C. (1998).
	\newblock Mutagen content and metabolic activation of promutagens by molluscs
	as biomarkers of marine pollution.
	\newblock \emph{Mutation Research/Fundamental and Molecular Mechanisms of
		Mutagenesis}, 399, 3--15.
	
	\bibitem[{Loreau(2010)}]{Loreau2010}
	\ifbool{MyRefNumbers}{\stepcounter{MyBibCount}\theMyBibCount.\\}{}Loreau, M.
	(2010).
	\newblock \emph{From populations to ecosystems: Theoretical foundations of a
		new ecological synthesis}.
	\newblock Princeton Universiry Press.
	
	\bibitem[{Loreau \emph{et~al.}(2003)Loreau, Mouquet \& Holt}]{Loreau2003}
	\ifbool{MyRefNumbers}{\stepcounter{MyBibCount}\theMyBibCount.\\}{}Loreau, M.,
	Mouquet, N. \& Holt, R.~D. (2003).
	\newblock Meta-ecosystems: a theoretical framework for a spatial ecosystem
	ecology.
	\newblock \emph{Ecol. Lett.}, 6, 673--679.
	
	\bibitem[{Louthan \emph{et~al.}(2015)Louthan, Doak \& Angert}]{Louthan2015}
	\ifbool{MyRefNumbers}{\stepcounter{MyBibCount}\theMyBibCount.\\}{}Louthan,
	A.~M., Doak, D.~F. \& Angert, A.~L. (2015).
	\newblock Where and when do species interactions set range limits?
	\newblock \emph{Trends Ecol. Evol.}, 30, 780--792.
	
	\bibitem[{Lynch(2007)}]{Lynch2007}
	\ifbool{MyRefNumbers}{\stepcounter{MyBibCount}\theMyBibCount.\\}{}Lynch, M.
	(2007).
	\newblock \emph{The Origin of Genome Architecture}.
	\newblock Sinauer Associates.
	
	\bibitem[{Lynch \emph{et~al.}(2016)Lynch, Ackerman, Gout, Long, Sung, Thomas \&
		Foster}]{Lynch2016}
	\ifbool{MyRefNumbers}{\stepcounter{MyBibCount}\theMyBibCount.\\}{}Lynch, M.,
	Ackerman, M.~S., Gout, J.-F., Long, H., Sung, W., Thomas, W.~K. \& Foster,
	P.~L. (2016).
	\newblock Genetic drift, selection and the evolution of the mutation rate.
	\newblock \emph{Nat. Rev. Genet.}, 17, 704--714.
	
	\bibitem[{Massol \emph{et~al.}(2011{\natexlab{a}})Massol, Duputi\'e, David \&
		Jarne}]{Massol2011}
	\ifbool{MyRefNumbers}{\stepcounter{MyBibCount}\theMyBibCount.\\}{}Massol, F.,
	Duputi\'e, A., David, P. \& Jarne, P. (2011{\natexlab{a}}).
	\newblock Asymmetric patch size distribution leads to disruptive selction on
	dispersal.
	\newblock \emph{Evolution}, 65, 490--500.
	
	\bibitem[{Massol \emph{et~al.}(2011{\natexlab{b}})Massol, Gravel, Mouquet,
		Cadotte, Fukami \& Leibold}]{Massol2011a}
	\ifbool{MyRefNumbers}{\stepcounter{MyBibCount}\theMyBibCount.\\}{}Massol, F.,
	Gravel, D., Mouquet, N., Cadotte, M.~W., Fukami, T. \& Leibold, M.~A.
	(2011{\natexlab{b}}).
	\newblock Linking community and ecosystem dynamics through spatial ecology.
	\newblock \emph{Ecol. Lett.}, 14, 313--323.
	
	\bibitem[{Masson-Delmotte \emph{et~al.}(2021)Masson-Delmotte, Zhai, Pirani,
		Connors, Péan, Berger, Caud, Chen, Goldfarb, Gomis, Huang, Leitzell, Lonnoy,
		Matthews, Maycock, Waterfield, Yelekçi, Yu \& Zhou}]{MassonDelmotte2021}
	\ifbool{MyRefNumbers}{\stepcounter{MyBibCount}\theMyBibCount.\\}{}Masson-Delmotte,
	V., Zhai, P., Pirani, A., Connors, S.~L., Péan, C., Berger, S., Caud, N.,
	Chen, Y., Goldfarb, L., Gomis, M.~I., Huang, M., Leitzell, K., Lonnoy, E.,
	Matthews, J. B.~R., Maycock, T.~K., Waterfield, T., Yelekçi, O., Yu, R. \&
	Zhou, B., eds. (2021).
	\newblock \emph{Climate {Change} 2021: {The} {Physical} {Science} {Basis}.
		{Contribution} of {Working} {Group} {I} to the {Sixth} {Assessment} {Report}
		of the {Intergovernmental} {Panel} on {Climate} {Change}}.
	\newblock Cambridge University Press.
	
	\bibitem[{Matthews \emph{et~al.}(2016)Matthews, Aebischer, Sullam,
		Lundsgaard-Hansen \& Seehausen}]{Matthews2016}
	\ifbool{MyRefNumbers}{\stepcounter{MyBibCount}\theMyBibCount.\\}{}Matthews, B.,
	Aebischer, T., Sullam, K.~E., Lundsgaard-Hansen, B. \& Seehausen, O. (2016).
	\newblock Experimental evidence of an eco-evolutionary feedback during adaptive
	divergence.
	\newblock \emph{Curr. Biol.}, 26, 483--489.
	
	\bibitem[{Matthews \emph{et~al.}(2011)Matthews, Narwani, Hausch, Nonaka, Peter,
		Yamamichi, Sullam, Bird, Thomas, Hanley \& Turner}]{Matthews2011}
	\ifbool{MyRefNumbers}{\stepcounter{MyBibCount}\theMyBibCount.\\}{}Matthews, B.,
	Narwani, A., Hausch, S., Nonaka, E., Peter, H., Yamamichi, M., Sullam, K.~E.,
	Bird, K.~C., Thomas, M.~K., Hanley, T.~C. \& Turner, C.~B. (2011).
	\newblock Toward an integration of evolutionary biology and ecosystem science.
	\newblock \emph{Ecol. Lett.}, 14, 690--701.
	
	\bibitem[{{Maynard Smith} \& Szathmáry(1995)}]{MaynardSmith1995}
	\ifbool{MyRefNumbers}{\stepcounter{MyBibCount}\theMyBibCount.\\}{}{Maynard
		Smith}, J. \& Szathmáry, E. (1995).
	\newblock \emph{The Major Transitions in Evolution}.
	\newblock Oxford University Press.
	
	\bibitem[{McPeek(2017)}]{McPeek2017a}
	\ifbool{MyRefNumbers}{\stepcounter{MyBibCount}\theMyBibCount.\\}{}McPeek, M.
	(2017).
	\newblock \emph{Evolutionary Community Ecology}.
	\newblock Princeton University Press.
	
	\bibitem[{McShea \& Brandon(2010)}]{McShea2010}
	\ifbool{MyRefNumbers}{\stepcounter{MyBibCount}\theMyBibCount.\\}{}McShea, D.~W.
	\& Brandon, R.~N. (2010).
	\newblock \emph{Biology’s First Law: The Tendency for Diversity and
		Complexity to Increase in Evolutionary Systems.}
	\newblock University of Chicago Press.
	
	\bibitem[{Meli\'an \emph{et~al.}(2018)Meli\'an, Matthews, de~Andreazzi,
		Rodr\'iguez, Harmon \& Fortuna}]{Melian2018}
	\ifbool{MyRefNumbers}{\stepcounter{MyBibCount}\theMyBibCount.\\}{}Meli\'an,
	C.~J., Matthews, B., de~Andreazzi, C.~S., Rodr\'iguez, J.~P., Harmon, L.~J.
	\& Fortuna, M.~A. (2018).
	\newblock Deciphering the interdependence between ecological and evolutionary
	networks.
	\newblock \emph{Trends Ecol. Evol.}, 33, 504--512.
	
	\bibitem[{Metz \& Gyllenberg(2001)}]{Metz2001}
	\ifbool{MyRefNumbers}{\stepcounter{MyBibCount}\theMyBibCount.\\}{}Metz, J.
	A.~J. \& Gyllenberg, M. (2001).
	\newblock How should we define fitness in structured metapopulation models?
	{Including} an application to the calculation of evolutionarily stable
	dispersal strategies.
	\newblock \emph{Proc. R. Soc. B-Biol. Sci.}, 268, 499--508.
	
	\bibitem[{Meysman \emph{et~al.}(2006)Meysman, Middelburg \& Heip}]{Meysman2006}
	\ifbool{MyRefNumbers}{\stepcounter{MyBibCount}\theMyBibCount.\\}{}Meysman, F.,
	Middelburg, J. \& Heip, C. (2006).
	\newblock Bioturbation: a fresh look at {D}arwin's last idea.
	\newblock \emph{Trends Ecol. Evol.}, 21, 688--695.
	
	\bibitem[{Miller \emph{et~al.}(2020)Miller, Angert, Brown, Lee-Yaw, Lewis,
		Lutscher, Marculis, Melbourne, Shaw, Sz{\H{u}}cs, Tabares, Usui, Weiss-Lehman
		\& Williams}]{Miller2020}
	\ifbool{MyRefNumbers}{\stepcounter{MyBibCount}\theMyBibCount.\\}{}Miller, T.
	E.~X., Angert, A.~L., Brown, C.~D., Lee-Yaw, J.~A., Lewis, M., Lutscher, F.,
	Marculis, N.~G., Melbourne, B.~A., Shaw, A.~K., Sz{\H{u}}cs, M., Tabares, O.,
	Usui, T., Weiss-Lehman, C. \& Williams, J.~L. (2020).
	\newblock Eco-evolutionary dynamics of range expansion.
	\newblock \emph{Ecology}, 101, e03139.
	
	\bibitem[{Mills \emph{et~al.}(2022)Mills, Boyle, Daines, Sperling, Pisani,
		Donoghue \& Lenton}]{Mills2022}
	\ifbool{MyRefNumbers}{\stepcounter{MyBibCount}\theMyBibCount.\\}{}Mills, D.~B.,
	Boyle, R.~A., Daines, S.~J., Sperling, E.~A., Pisani, D., Donoghue, P. C.~J.
	\& Lenton, T.~M. (2022).
	\newblock Eukaryogenesis and oxygen in earth history.
	\newblock \emph{Nat. Ecol. Evol.}, 6, 520--532.
	
	\bibitem[{Moerman \emph{et~al.}(2020)Moerman, Fronhofer, Wagner \&
		Altermatt}]{Moerman2020a}
	\ifbool{MyRefNumbers}{\stepcounter{MyBibCount}\theMyBibCount.\\}{}Moerman, F.,
	Fronhofer, E.~A., Wagner, A. \& Altermatt, F. (2020).
	\newblock Sex and gene flow modulate evolution during range expansions in the
	protist \textit{Tetrahymena thermophila}.
	\newblock \emph{Biol. Lett.}, 16, 20200244.
	
	\bibitem[{Morgan \emph{et~al.}(2017)Morgan, Zhang \& Bomblies}]{Morgan2017}
	\ifbool{MyRefNumbers}{\stepcounter{MyBibCount}\theMyBibCount.\\}{}Morgan,
	C.~H., Zhang, H. \& Bomblies, K. (2017).
	\newblock Are the effects of elevated temperature on meiotic recombination and
	thermotolerance linked via the axis and synaptonemal complex?
	\newblock \emph{Philos. Trans. R. Soc. B-Biol. Sci.}, 372, 20160470.
	
	\bibitem[{Mouquet \emph{et~al.}(2015)Mouquet, Lagadeuc, Devictor, Doyen,
		Duputi{\'{e}}, Eveillard, Faure, Garnier, Gimenez, Huneman, Jabot, Jarne,
		Joly, Julliard, K{\'{e}}fi, Kergoat, Lavorel, Gall, Meslin, Morand, Morin,
		Morlon, Pinay, Pradel, Schurr, Thuiller \& Loreau}]{Mouquet2015}
	\ifbool{MyRefNumbers}{\stepcounter{MyBibCount}\theMyBibCount.\\}{}Mouquet, N.,
	Lagadeuc, Y., Devictor, V., Doyen, L., Duputi{\'{e}}, A., Eveillard, D.,
	Faure, D., Garnier, E., Gimenez, O., Huneman, P., Jabot, F., Jarne, P., Joly,
	D., Julliard, R., K{\'{e}}fi, S., Kergoat, G.~J., Lavorel, S., Gall, L.~L.,
	Meslin, L., Morand, S., Morin, X., Morlon, H., Pinay, G., Pradel, R., Schurr,
	F.~M., Thuiller, W. \& Loreau, M. (2015).
	\newblock Predictive ecology in a changing world.
	\newblock \emph{J. Appl. Ecol.}, 52, 1293--1310.
	
	\bibitem[{Murray \emph{et~al.}(2008)Murray, Knaapen, Tal \&
		Kirwan}]{Murray2008}
	\ifbool{MyRefNumbers}{\stepcounter{MyBibCount}\theMyBibCount.\\}{}Murray,
	A.~B., Knaapen, M. A.~F., Tal, M. \& Kirwan, M.~L. (2008).
	\newblock Biomorphodynamics: Physical-biological feedbacks that shape
	landscapes.
	\newblock \emph{Water Resour Res}, 44, W11301.
	
	\bibitem[{Nichol \emph{et~al.}(2019)Nichol, Robertson-Tessi, Anderson \&
		Jeavons}]{Nichol2019}
	\ifbool{MyRefNumbers}{\stepcounter{MyBibCount}\theMyBibCount.\\}{}Nichol, D.,
	Robertson-Tessi, M., Anderson, A.~R. \& Jeavons, P. (2019).
	\newblock Model genotype--phenotype mappings and the algorithmic structure of
	evolution.
	\newblock \emph{J. R. Soc. Interface}, 16, 20190332.
	
	\bibitem[{Nikas \emph{et~al.}(1985)Nikas, Tiffney \& Knoll.}]{Nikas1985}
	\ifbool{MyRefNumbers}{\stepcounter{MyBibCount}\theMyBibCount.\\}{}Nikas, K.~J.,
	Tiffney, B.~H. \& Knoll., A.~H. (1985).
	\newblock Patterns in vascular land plant diversification: An analysis at the
	species level.
	\newblock In: \emph{Phanerozoic Diversity Patterns} (ed. Valentine, J.).
	
	\bibitem[{Odling-Smee \emph{et~al.}(2003)Odling-Smee, Odling-Smee, Laland,
		Feldman \& Feldman}]{Odling-Smee2003}
	\ifbool{MyRefNumbers}{\stepcounter{MyBibCount}\theMyBibCount.\\}{}Odling-Smee,
	F.~J., Odling-Smee, H., Laland, K.~N., Feldman, M.~W. \& Feldman, F. (2003).
	\newblock \emph{Niche construction: the neglected process in evolution}.
	\newblock Princeton University Press.
	
	\bibitem[{Oziolor \emph{et~al.}(2019)Oziolor, Reid, Yair, Lee, VerPloeg, Bruns,
		Shaw, Whitehead \& Matson}]{Oziolor2019}
	\ifbool{MyRefNumbers}{\stepcounter{MyBibCount}\theMyBibCount.\\}{}Oziolor,
	E.~M., Reid, N.~M., Yair, S., Lee, K.~M., VerPloeg, S.~G., Bruns, P.~C.,
	Shaw, J.~R., Whitehead, A. \& Matson, C.~W. (2019).
	\newblock Adaptive introgression enables evolutionary rescue from extreme
	environmental pollution.
	\newblock \emph{Science}, 364, 455--457.
	
	\bibitem[{Pardo \emph{et~al.}(2017)Pardo, Forcada, Wood, Tuck, Ireland, Pradel,
		Croxall \& Phillips}]{Pardo2017}
	\ifbool{MyRefNumbers}{\stepcounter{MyBibCount}\theMyBibCount.\\}{}Pardo, D.,
	Forcada, J., Wood, A.~G., Tuck, G.~N., Ireland, L., Pradel, R., Croxall,
	J.~P. \& Phillips, R.~A. (2017).
	\newblock Additive effects of climate and fisheries drive ongoing declines in
	multiple albatross species.
	\newblock \emph{Proc. Natl. Acad. Sci. U. S. A.}, 114, E10829--E10837.
	
	\bibitem[{Pausas \& Bond(2022)}]{Pausas2022}
	\ifbool{MyRefNumbers}{\stepcounter{MyBibCount}\theMyBibCount.\\}{}Pausas, J.~G.
	\& Bond, W.~J. (2022).
	\newblock Feedbacks in ecology and evolution.
	\newblock \emph{Trends Ecol. Evol.}, 37, 637--644.
	
	\bibitem[{Payne \& Wagner(2019)}]{Payne2019}
	\ifbool{MyRefNumbers}{\stepcounter{MyBibCount}\theMyBibCount.\\}{}Payne, J.~L.
	\& Wagner, A. (2019).
	\newblock The causes of evolvability and their evolution.
	\newblock \emph{Nat. Rev. Genet.}, 20, 24--38.
	
	\bibitem[{Pelletier \emph{et~al.}(2009)Pelletier, Garant \&
		Hendry}]{Pelletier2009}
	\ifbool{MyRefNumbers}{\stepcounter{MyBibCount}\theMyBibCount.\\}{}Pelletier,
	F., Garant, D. \& Hendry, A. (2009).
	\newblock Eco-evolutionary dynamics.
	\newblock \emph{Philos. Trans. R. Soc. B-Biol. Sci.}, 364, 1483--1489.
	
	\bibitem[{Penny \& Phillips(2004)}]{Penny2004}
	\ifbool{MyRefNumbers}{\stepcounter{MyBibCount}\theMyBibCount.\\}{}Penny, D. \&
	Phillips, M.~J. (2004).
	\newblock The rise of birds and mammals: are microevolutionary processes
	sufficient for macroevolution?
	\newblock \emph{Trends Ecol. Evol.}, 19, 516--522.
	
	\bibitem[{Perkins-Kirkpatrick \& Lewis(2020)}]{PerkinsKirkpatrick2020}
	\ifbool{MyRefNumbers}{\stepcounter{MyBibCount}\theMyBibCount.\\}{}Perkins-Kirkpatrick,
	S.~E. \& Lewis, S.~C. (2020).
	\newblock Increasing trends in regional heatwaves.
	\newblock \emph{Nat. Commun.}, 11.
	
	\bibitem[{Peters \emph{et~al.}(2014)Peters, Havstad, Cushing, Tweedie, Fuentes
		\& Villanueva-Rosales}]{Peters2014}
	\ifbool{MyRefNumbers}{\stepcounter{MyBibCount}\theMyBibCount.\\}{}Peters, D.
	P.~C., Havstad, K.~M., Cushing, J., Tweedie, C., Fuentes, O. \&
	Villanueva-Rosales, N. (2014).
	\newblock Harnessing the power of big data: infusing the scientific method with
	machine learning to transform ecology.
	\newblock \emph{Ecosphere}, 5, art67.
	
	\bibitem[{Phillips(2021)}]{Phillips2021}
	\ifbool{MyRefNumbers}{\stepcounter{MyBibCount}\theMyBibCount.\\}{}Phillips, J.
	(2021).
	\newblock \emph{Landscape Evolution: Landforms, Ecosystems, and Soils}.
	\newblock Elsevier.
	
	\bibitem[{Phillips(2015)}]{Phillips2015}
	\ifbool{MyRefNumbers}{\stepcounter{MyBibCount}\theMyBibCount.\\}{}Phillips,
	J.~D. (2015).
	\newblock Landforms as extended composite phenotypes.
	\newblock \emph{Earth Surf Proc Land}, 41, 16--26.
	
	\bibitem[{Pigliucci(2008)}]{Pigliucci2008a}
	\ifbool{MyRefNumbers}{\stepcounter{MyBibCount}\theMyBibCount.\\}{}Pigliucci, M.
	(2008).
	\newblock Is evolvability evolvable?
	\newblock \emph{Nature Reviews Genetics}, 9, 75--82.
	
	\bibitem[{Pilosof \emph{et~al.}(2017)Pilosof, Porter, Pascual \&
		K{\'e}fi}]{Pilosof2017}
	\ifbool{MyRefNumbers}{\stepcounter{MyBibCount}\theMyBibCount.\\}{}Pilosof, S.,
	Porter, M.~A., Pascual, M. \& K{\'e}fi, S. (2017).
	\newblock The multilayer nature of ecological networks.
	\newblock \emph{Nat. Ecol. Evol.}, 1, 0101.
	
	\bibitem[{Pimentel(1961)}]{Pimentel1961}
	\ifbool{MyRefNumbers}{\stepcounter{MyBibCount}\theMyBibCount.\\}{}Pimentel, D.
	(1961).
	\newblock Animal population regulation by the genetic {Feed-Back} mechanism.
	\newblock \emph{Am. Nat.}, 95, 65--79.
	
	\bibitem[{Pimentel(1968)}]{Pimentel1968}
	\ifbool{MyRefNumbers}{\stepcounter{MyBibCount}\theMyBibCount.\\}{}Pimentel, D.
	(1968).
	\newblock Population regulation and genetic feedback. evolution provides
	foundation for control of herbivore, parasite, and predator numbers in
	nature.
	\newblock \emph{Science}, 159, 1432--1437.
	
	\bibitem[{Poethke \& Hovestadt(2002)}]{Poethke2002}
	\ifbool{MyRefNumbers}{\stepcounter{MyBibCount}\theMyBibCount.\\}{}Poethke,
	H.~J. \& Hovestadt, T. (2002).
	\newblock Evolution of density- and patch-size-dependent dispersal rates.
	\newblock \emph{Proc. R. Soc. B-Biol. Sci.}, 269, 637--645.
	
	\bibitem[{Post \& Palkovacs(2009)}]{Post2009}
	\ifbool{MyRefNumbers}{\stepcounter{MyBibCount}\theMyBibCount.\\}{}Post, D.~M.
	\& Palkovacs, E.~P. (2009).
	\newblock Eco-evolutionary feedbacks in community and ecosystem ecology:
	interactions between the ecological theatre and the evolutionary play.
	\newblock \emph{Philos. Trans. R. Soc. B-Biol. Sci.}, 364, 1629--1640.
	
	\bibitem[{Rabosky \emph{et~al.}(2018)Rabosky, Chang, Title, Cowman, Sallan,
		Friedman, Kaschner, Garilao, Near, Coll \& Alfaro}]{Rabosky2018}
	\ifbool{MyRefNumbers}{\stepcounter{MyBibCount}\theMyBibCount.\\}{}Rabosky,
	D.~L., Chang, J., Title, P.~O., Cowman, P.~F., Sallan, L., Friedman, M.,
	Kaschner, K., Garilao, C., Near, T.~J., Coll, M. \& Alfaro, M.~E. (2018).
	\newblock An inverse latitudinal gradient in speciation rate for marine fishes.
	\newblock \emph{Nature}, 559, 392--395.
	
	\bibitem[{Rammer \& Seidl(2019)}]{Rammer2019}
	\ifbool{MyRefNumbers}{\stepcounter{MyBibCount}\theMyBibCount.\\}{}Rammer, W. \&
	Seidl, R. (2019).
	\newblock Harnessing deep learning in ecology: An example predicting bark
	beetle outbreaks.
	\newblock \emph{Front. Plant Sci.}, 10, 1327.
	
	\bibitem[{Raup \& Sepkoski(1982)}]{Raup1982}
	\ifbool{MyRefNumbers}{\stepcounter{MyBibCount}\theMyBibCount.\\}{}Raup, D.~M.
	\& Sepkoski, J.~J. (1982).
	\newblock Mass extinctions in the marine fossil record.
	\newblock \emph{Science}, 215, 1501--1503.
	
	\bibitem[{Reid \emph{et~al.}(2016)Reid, Proestou, Clark, Warren, Colbourne,
		Shaw, Karchner, Hahn, Nacci, Oleksiak, Crawford \& Whitehead}]{Reid2016}
	\ifbool{MyRefNumbers}{\stepcounter{MyBibCount}\theMyBibCount.\\}{}Reid, N.~M.,
	Proestou, D.~A., Clark, B.~W., Warren, W.~C., Colbourne, J.~K., Shaw, J.~R.,
	Karchner, S.~I., Hahn, M.~E., Nacci, D., Oleksiak, M.~F., Crawford, D.~L. \&
	Whitehead, A. (2016).
	\newblock The genomic landscape of rapid repeated evolutionary adaptation to
	toxic pollution in wild fish.
	\newblock \emph{Science}, 354, 1305--1308.
	
	\bibitem[{Reznick(1982)}]{Reznick1982}
	\ifbool{MyRefNumbers}{\stepcounter{MyBibCount}\theMyBibCount.\\}{}Reznick, D.
	(1982).
	\newblock The impact of predation on life history evolution in trinidadian
	guppies: Genetic basis of observed life history patterns.
	\newblock \emph{Evolution}, 36, 1236--1250.
	
	\bibitem[{Reznick \& Travis(2019)}]{Reznick2019}
	\ifbool{MyRefNumbers}{\stepcounter{MyBibCount}\theMyBibCount.\\}{}Reznick,
	D.~N. \& Travis, J. (2019).
	\newblock Experimental studies of evolution and eco-evo dynamics in guppies
	(\textit{Poecilia reticulata}).
	\newblock \emph{Annu. Rev. Ecol. Evol. Syst.}, 50, 335--354.
	
	\bibitem[{Riederer \emph{et~al.}(2022)Riederer, Tiso, van Eldijk \&
		J.Weissing}]{Riederer2022}
	\ifbool{MyRefNumbers}{\stepcounter{MyBibCount}\theMyBibCount.\\}{}Riederer,
	J.~M., Tiso, S., van Eldijk, T. J.~B. \& J.Weissing, F. (2022).
	\newblock Capturing the facets of evolvability in a mechanistic framework.
	\newblock \emph{Trends Ecol. Evol.}, 37, 430--439.
	
	\bibitem[{Romiguier \emph{et~al.}(2014)Romiguier, Gayral, Ballenghien, Bernard,
		Cahais, Chenuil, Chiari, Dernat, Duret, Faivre, Loire, Lourenco, Nabholz,
		Roux, Tsagkogeorga, Weber, Weinert, Belkhir, Bierne, Gl{\'{e}}min \&
		Galtier}]{Romiguier2014}
	\ifbool{MyRefNumbers}{\stepcounter{MyBibCount}\theMyBibCount.\\}{}Romiguier,
	J., Gayral, P., Ballenghien, M., Bernard, A., Cahais, V., Chenuil, A.,
	Chiari, Y., Dernat, R., Duret, L., Faivre, N., Loire, E., Lourenco, J.~M.,
	Nabholz, B., Roux, C., Tsagkogeorga, G., Weber, A. A.-T., Weinert, L.~A.,
	Belkhir, K., Bierne, N., Gl{\'{e}}min, S. \& Galtier, N. (2014).
	\newblock Comparative population genomics in animals uncovers the determinants
	of genetic diversity.
	\newblock \emph{Nature}, 515, 261--263.
	
	\bibitem[{Ronce(2007)}]{Ronce2007}
	\ifbool{MyRefNumbers}{\stepcounter{MyBibCount}\theMyBibCount.\\}{}Ronce, O.
	(2007).
	\newblock How does it feel to be like a rolling stone? {Ten} questions about
	dispersal evolution.
	\newblock \emph{Annu. Rev. Ecol. Evol. Syst.}, 38, 231--253.
	
	\bibitem[{Rose(1991)}]{Rose1991}
	\ifbool{MyRefNumbers}{\stepcounter{MyBibCount}\theMyBibCount.\\}{}Rose, M.
	(1991).
	\newblock \emph{Evolutionary Biology of Aging}.
	\newblock Oxford University Press.
	
	\bibitem[{Rudman \emph{et~al.}(2017)Rudman, Barbour, Csill{\'{e}}ry, Gienapp,
		Guillaume, Jr, Hendry, Lasky, Rafajlovi{\'{c}}, Räsänen, Schmidt,
		Seehausen, Therkildsen, Turcotte \& Levine}]{Rudman2017}
	\ifbool{MyRefNumbers}{\stepcounter{MyBibCount}\theMyBibCount.\\}{}Rudman,
	S.~M., Barbour, M.~A., Csill{\'{e}}ry, K., Gienapp, P., Guillaume, F., Jr, N.
	G.~H., Hendry, A.~P., Lasky, J.~R., Rafajlovi{\'{c}}, M., Räsänen, K.,
	Schmidt, P.~S., Seehausen, O., Therkildsen, N.~O., Turcotte, M.~M. \& Levine,
	J.~M. (2017).
	\newblock What genomic data can reveal about eco-evolutionary dynamics.
	\newblock \emph{Nat Ecol Evol}, 2, 9--15.
	
	\bibitem[{Saaristo \emph{et~al.}(2018)Saaristo, Brodin, Balshine, Bertram,
		Brooks, Ehlman, McCallum, Sih, Sundin, Wong \& Arnold}]{Saaristo2018}
	\ifbool{MyRefNumbers}{\stepcounter{MyBibCount}\theMyBibCount.\\}{}Saaristo, M.,
	Brodin, T., Balshine, S., Bertram, M.~G., Brooks, B.~W., Ehlman, S.~M.,
	McCallum, E.~S., Sih, A., Sundin, J., Wong, B. B.~M. \& Arnold, K.~E. (2018).
	\newblock Direct and indirect effects of chemical contaminants on the
	behaviour, ecology and evolution of wildlife.
	\newblock \emph{Proc. R. Soc. B-Biol. Sci.}, 285, 20181297.
	
	\bibitem[{Saastamoinen \emph{et~al.}(2018)Saastamoinen, Bocedi, Cote, Legrand,
		Guillaume, Wheat, Fronhofer, Garcia, Henry, Husby, Baguette, Bonte, Coulon,
		Kokko, Matthysen, Niitep{\~o}ld, Nonaka, Stevens, Travis, Donohue, Bullock \&
		del Mar~Delgado}]{Saastamoinen2018}
	\ifbool{MyRefNumbers}{\stepcounter{MyBibCount}\theMyBibCount.\\}{}Saastamoinen,
	M., Bocedi, G., Cote, J., Legrand, D., Guillaume, F., Wheat, C.~W.,
	Fronhofer, E.~A., Garcia, C., Henry, R., Husby, A., Baguette, M., Bonte, D.,
	Coulon, A., Kokko, H., Matthysen, E., Niitep{\~o}ld, K., Nonaka, E., Stevens,
	V.~M., Travis, J. M.~J., Donohue, K., Bullock, J.~M. \& del Mar~Delgado, M.
	(2018).
	\newblock Genetics of dispersal.
	\newblock \emph{Biol. Rev.}, 93, 574--599.
	
	\bibitem[{Scheffer \emph{et~al.}(2001)Scheffer, Carpenter, Foley, Folke \&
		Walker}]{Scheffer2001}
	\ifbool{MyRefNumbers}{\stepcounter{MyBibCount}\theMyBibCount.\\}{}Scheffer, M.,
	Carpenter, S., Foley, J.~A., Folke, C. \& Walker, B. (2001).
	\newblock Catastrophic shifts in ecosystems.
	\newblock \emph{Nature}, 413, 591--596.
	
	\bibitem[{Schluter(1996)}]{Schluter1996}
	\ifbool{MyRefNumbers}{\stepcounter{MyBibCount}\theMyBibCount.\\}{}Schluter, D.
	(1996).
	\newblock Adaptive radiation along genetic lines of least resistance.
	\newblock \emph{Evolution}, 50, 1766--1774.
	
	\bibitem[{Shoemaker \emph{et~al.}(2020)Shoemaker, Sullivan, Donohue, Cabral,
		Williams, Mayfield, Chase, Chu, Harpole, Huth, HilleRisLambers, James, Kraft,
		May, Muthukrishnan, Satterlee, Taubert, Wang, Wiegand, Yang \&
		Abbott}]{Shoemaker2020}
	\ifbool{MyRefNumbers}{\stepcounter{MyBibCount}\theMyBibCount.\\}{}Shoemaker,
	L.~G., Sullivan, L.~L., Donohue, I., Cabral, J.~S., Williams, R.~J.,
	Mayfield, M.~M., Chase, J.~M., Chu, C., Harpole, W.~S., Huth, A.,
	HilleRisLambers, J., James, A. R.~M., Kraft, N. J.~B., May, F.,
	Muthukrishnan, R., Satterlee, S., Taubert, F., Wang, X., Wiegand, T., Yang,
	Q. \& Abbott, K.~C. (2020).
	\newblock Integrating the underlying structure of stochasticity into community
	ecology.
	\newblock \emph{Ecology}, 101, e02922.
	
	\bibitem[{Siepielski \emph{et~al.}(2020)Siepielski, Hasik, Ping, Serrano,
		Strayhorn \& Tye}]{Siepielski2020}
	\ifbool{MyRefNumbers}{\stepcounter{MyBibCount}\theMyBibCount.\\}{}Siepielski,
	A.~M., Hasik, A.~Z., Ping, T., Serrano, M., Strayhorn, K. \& Tye, S.~P.
	(2020).
	\newblock Predators weaken prey intraspecific competition through phenotypic
	selection.
	\newblock \emph{Ecol. Lett.}, 23, 951--961.
	
	\bibitem[{Slobodkin(1961)}]{Slobodkin1961}
	\ifbool{MyRefNumbers}{\stepcounter{MyBibCount}\theMyBibCount.\\}{}Slobodkin,
	L.~B. (1961).
	\newblock \emph{Growth and Regulation of Animal Populations}.
	\newblock Holt, Rinehart and Winston.
	
	\bibitem[{Sniegowski \emph{et~al.}(2000)Sniegowski, Gerrish, Johnson \&
		Shaver}]{Sniegowski2000}
	\ifbool{MyRefNumbers}{\stepcounter{MyBibCount}\theMyBibCount.\\}{}Sniegowski,
	P.~D., Gerrish, P.~J., Johnson, T. \& Shaver, A. (2000).
	\newblock The evolution of mutation rates: separating causes from consequences.
	\newblock \emph{BioEssays}, 22, 1057--1066.
	
	\bibitem[{Sol{\'{e}} \& Levin(2022)}]{Sole2022}
	\ifbool{MyRefNumbers}{\stepcounter{MyBibCount}\theMyBibCount.\\}{}Sol{\'{e}},
	R. \& Levin, S. (2022).
	\newblock Ecological complexity and the biosphere: the next 30 years.
	\newblock \emph{Philos. Trans. R. Soc. B-Biol. Sci.}, 377.
	
	\bibitem[{Somers \emph{et~al.}(2002)Somers, Yauk, White, Parfett \&
		Quinn}]{Somers2002}
	\ifbool{MyRefNumbers}{\stepcounter{MyBibCount}\theMyBibCount.\\}{}Somers,
	C.~M., Yauk, C.~L., White, P.~A., Parfett, C. L.~J. \& Quinn, J.~S. (2002).
	\newblock Air pollution induces heritable {DNA} mutations.
	\newblock \emph{Proc. Natl. Acad. Sci. U. S. A.}, 99, 15904--15907.
	
	\bibitem[{Stoks \emph{et~al.}(2015)Stoks, Govaert, Pauwels, Jansen \&
		De~Meester}]{Stoks2015}
	\ifbool{MyRefNumbers}{\stepcounter{MyBibCount}\theMyBibCount.\\}{}Stoks, R.,
	Govaert, L., Pauwels, K., Jansen, B. \& De~Meester, L. (2015).
	\newblock Resurrecting complexity: the interplay of plasticity and rapid
	evolution in the multiple trait response to strong changes in predation
	pressure in the water flea daphnia magna.
	\newblock \emph{Ecol. Lett.}, 19, 180--190.
	
	\bibitem[{Stroud \& Losos(2016)}]{Stroud2016}
	\ifbool{MyRefNumbers}{\stepcounter{MyBibCount}\theMyBibCount.\\}{}Stroud, J.~T.
	\& Losos, J.~B. (2016).
	\newblock Ecological opportunity and adaptive radiation.
	\newblock \emph{Annu. Rev. Ecol. Evol. Syst.}, 47.
	
	\bibitem[{Synodinos \emph{et~al.}(2021)Synodinos, Haegeman, Sentis \&
		Montoya}]{Synodinos2021}
	\ifbool{MyRefNumbers}{\stepcounter{MyBibCount}\theMyBibCount.\\}{}Synodinos,
	A.~D., Haegeman, B., Sentis, A. \& Montoya, J.~M. (2021).
	\newblock Theory of temperature-dependent consumer-resource interactions.
	\newblock \emph{Ecology Letters}, 24, 1539--1555.
	
	\bibitem[{Szathm{\'{a}}ry \& {Maynard Smith}(1995)}]{Szathmary1995}
	\ifbool{MyRefNumbers}{\stepcounter{MyBibCount}\theMyBibCount.\\}{}Szathm{\'{a}}ry,
	E. \& {Maynard Smith}, J. (1995).
	\newblock The major evolutionary transitions.
	\newblock \emph{Nature}, 374, 227--232.
	
	\bibitem[{Urban(2007)}]{Urban2007}
	\ifbool{MyRefNumbers}{\stepcounter{MyBibCount}\theMyBibCount.\\}{}Urban, M.
	(2007).
	\newblock Risky prey behavior evolves in risky habitats.
	\newblock \emph{Proc. Natl. Acad. Sci. U. S. A.}, 104, 14377--14382.
	
	\bibitem[{Urban \emph{et~al.}(2016)Urban, Bocedi, Hendry, Mihoub, Peer, Singer,
		Bridle, Crozier, Meester, Godsoe, Gonzalez, Hellmann, Holt, Huth, Johst,
		Krug, Leadley, Palmer, Pantel, Schmitz, Zollner \& Travis}]{Urban2016}
	\ifbool{MyRefNumbers}{\stepcounter{MyBibCount}\theMyBibCount.\\}{}Urban, M.~C.,
	Bocedi, G., Hendry, A.~P., Mihoub, J.-B., Peer, G., Singer, A., Bridle,
	J.~R., Crozier, L.~G., Meester, L.~D., Godsoe, W., Gonzalez, A., Hellmann,
	J.~J., Holt, R.~D., Huth, A., Johst, K., Krug, C.~B., Leadley, P.~W., Palmer,
	S. C.~F., Pantel, J.~H., Schmitz, A., Zollner, P.~A. \& Travis, J. M.~J.
	(2016).
	\newblock Improving the forecast for biodiversity under climate change.
	\newblock \emph{Science}, 353, aad8466--1 -- aad8466--9.
	
	\bibitem[{Van~Gestel \& Weissing(2016)}]{VanGestel2016}
	\ifbool{MyRefNumbers}{\stepcounter{MyBibCount}\theMyBibCount.\\}{}Van~Gestel,
	J. \& Weissing, F.~J. (2016).
	\newblock Regulatory mechanisms link phenotypic plasticity to evolvability.
	\newblock \emph{Sci. Rep.}, 6, 24524.
	
	\bibitem[{Vermeij(2017)}]{Vermeij2017}
	\ifbool{MyRefNumbers}{\stepcounter{MyBibCount}\theMyBibCount.\\}{}Vermeij,
	G.~J. (2017).
	\newblock How the land became the locus of major evolutionary innovations.
	\newblock \emph{Curr. Biol.}, 27, 3178--3182.e1.
	
	\bibitem[{Wagner(2011)}]{Wagner2011}
	\ifbool{MyRefNumbers}{\stepcounter{MyBibCount}\theMyBibCount.\\}{}Wagner, A.
	(2011).
	\newblock \emph{The origins of evolutionary innovations: A theory of
		transformative change in living systems}.
	\newblock Oxford University Press.
	
	\bibitem[{Wagner \& Altenberg(1996)}]{Wagner1996}
	\ifbool{MyRefNumbers}{\stepcounter{MyBibCount}\theMyBibCount.\\}{}Wagner, G.~P.
	\& Altenberg, L. (1996).
	\newblock Perspective: Complex adaptations and the evolution of evolvability.
	\newblock \emph{Evolution}, 50, 967--976.
	
	\bibitem[{Waldvogel \& Pfenninger(2021)}]{Waldvogel2021}
	\ifbool{MyRefNumbers}{\stepcounter{MyBibCount}\theMyBibCount.\\}{}Waldvogel,
	A.-M. \& Pfenninger, M. (2021).
	\newblock Temperature dependence of spontaneous mutation rates.
	\newblock \emph{Genome Res.}, 31, 1582--1589.
	
	\bibitem[{Weir \& Schluter(2007)}]{Weir2007}
	\ifbool{MyRefNumbers}{\stepcounter{MyBibCount}\theMyBibCount.\\}{}Weir, J.~T.
	\& Schluter, D. (2007).
	\newblock The latitudinal gradient in recent speciation and extinction rates of
	birds and mammals.
	\newblock \emph{Science}, 315, 1574--1576.
	
	\bibitem[{Whitham \emph{et~al.}(2006)Whitham, Bailey, Schweitzer, Shuster,
		Bangert, LeRoy, Lonsdorf, Allan, DiFazio, Potts, Fischer, Gehring, Lindroth,
		Marks, Hart, Wimp \& Wooley}]{Whitham2006}
	\ifbool{MyRefNumbers}{\stepcounter{MyBibCount}\theMyBibCount.\\}{}Whitham,
	T.~G., Bailey, J.~K., Schweitzer, J.~A., Shuster, S.~M., Bangert, R.~K.,
	LeRoy, C.~J., Lonsdorf, E.~V., Allan, G.~J., DiFazio, S.~P., Potts, B.~M.,
	Fischer, D.~G., Gehring, C.~A., Lindroth, R.~L., Marks, J.~C., Hart, S.~C.,
	Wimp, G.~M. \& Wooley, S.~C. (2006).
	\newblock A framework for community and ecosystem genetics: from genes to
	ecosystems.
	\newblock \emph{Nat. Rev. Genet.}, 7, 510--523.
	
	\bibitem[{Wielgoss \emph{et~al.}(2012)Wielgoss, Barrick, Tenaillon, Wiser,
		Dittmar, Cruveiller, Chane-Woon-Ming, M{\'{e}}digue, Lenski \&
		Schneider}]{Wielgoss2012}
	\ifbool{MyRefNumbers}{\stepcounter{MyBibCount}\theMyBibCount.\\}{}Wielgoss, S.,
	Barrick, J.~E., Tenaillon, O., Wiser, M.~J., Dittmar, W.~J., Cruveiller, S.,
	Chane-Woon-Ming, B., M{\'{e}}digue, C., Lenski, R.~E. \& Schneider, D.
	(2012).
	\newblock Mutation rate dynamics in a bacterial population reflect tension
	between adaptation and genetic load.
	\newblock \emph{Proc. Natl. Acad. Sci. U. S. A.}, 110, 222--227.
	
	\bibitem[{Wilson \emph{et~al.}(2020)Wilson, White, Monta{\~{n}}ez, DiMichele,
		McElwain, Poulsen \& Hren}]{Wilson2020}
	\ifbool{MyRefNumbers}{\stepcounter{MyBibCount}\theMyBibCount.\\}{}Wilson,
	J.~P., White, J.~D., Monta{\~{n}}ez, I.~P., DiMichele, W.~A., McElwain,
	J.~C., Poulsen, C.~J. \& Hren, M.~T. (2020).
	\newblock Carboniferous plant physiology breaks the mold.
	\newblock \emph{New Phytol.}, 227, 667--679.
	
	\bibitem[{Yamamichi(2022)}]{Yamamichi2022}
	\ifbool{MyRefNumbers}{\stepcounter{MyBibCount}\theMyBibCount.\\}{}Yamamichi, M.
	(2022).
	\newblock How does genetic architecture affect eco-evolutionary dynamics? a
	theoretical perspective.
	\newblock \emph{Philos. Trans. R. Soc. B-Biol. Sci.}, 377.
	
	\bibitem[{Yates \emph{et~al.}(2018)Yates, Bouchet, Caley, Mengersen, Randin,
		Parnell, Fielding, Bamford, Ban, Barbosa, Dormann, Elith, Embling, Ervin,
		Fisher, Gould, Graf, Gregr, Halpin, Heikkinen, Heinänen, Jones,
		Krishnakumar, Lauria, Lozano-Montes, Mannocci, Mellin, Mesgaran, Moreno-Amat,
		Mormede, Novaczek, Oppel, Crespo, Peterson, Rapacciuolo, Roberts, Ross,
		Scales, Schoeman, Snelgrove, Sundblad, Thuiller, Torres, Verbruggen, Wang,
		Wenger, Whittingham, Zharikov, Zurell \& Sequeira}]{Yates2018}
	\ifbool{MyRefNumbers}{\stepcounter{MyBibCount}\theMyBibCount.\\}{}Yates, K.~L.,
	Bouchet, P.~J., Caley, M.~J., Mengersen, K., Randin, C.~F., Parnell, S.,
	Fielding, A.~H., Bamford, A.~J., Ban, S., Barbosa, A.~M., Dormann, C.~F.,
	Elith, J., Embling, C.~B., Ervin, G.~N., Fisher, R., Gould, S., Graf, R.~F.,
	Gregr, E.~J., Halpin, P.~N., Heikkinen, R.~K., Heinänen, S., Jones, A.~R.,
	Krishnakumar, P.~K., Lauria, V., Lozano-Montes, H., Mannocci, L., Mellin, C.,
	Mesgaran, M.~B., Moreno-Amat, E., Mormede, S., Novaczek, E., Oppel, S.,
	Crespo, G.~O., Peterson, A.~T., Rapacciuolo, G., Roberts, J.~J., Ross, R.~E.,
	Scales, K.~L., Schoeman, D., Snelgrove, P., Sundblad, G., Thuiller, W.,
	Torres, L.~G., Verbruggen, H., Wang, L., Wenger, S., Whittingham, M.~J.,
	Zharikov, Y., Zurell, D. \& Sequeira, A.~M. (2018).
	\newblock Outstanding challenges in the transferability of ecological models.
	\newblock \emph{Trends Ecol. Evol.}, 33, 790--802.
	
	\bibitem[{Yoshida \emph{et~al.}(2003)Yoshida, Jones, Ellner, Fussmann \&
		Hairston}]{Yoshida2003}
	\ifbool{MyRefNumbers}{\stepcounter{MyBibCount}\theMyBibCount.\\}{}Yoshida, T.,
	Jones, L.~E., Ellner, S.~P., Fussmann, G.~F. \& Hairston, N.~G. (2003).
	\newblock Rapid evolution drives ecological dynamics in a predator–prey
	system.
	\newblock \emph{Nature}, 424, 303--306.
	
\end{thebibliography}
\end{document}